\documentclass[12pt]{article}
\usepackage{graphicx}
\usepackage{epstopdf}
\usepackage{amsmath}
\usepackage{amsfonts}
\usepackage{amssymb}
\usepackage{latexsym}
\usepackage{hyperref}
\usepackage{bm}
\usepackage{bbold}
\usepackage[english]{babel}
\usepackage[utf8]{inputenc}
\usepackage{xcolor}
\usepackage{graphicx}
\usepackage{caption}
\usepackage[caption=false]{subfig}
\usepackage{cite}
\usepackage{float}

\usepackage{graphicx}
\usepackage[caption=false]{subfig}
\usepackage{epstopdf}
\usepackage{amsmath}
\usepackage{amsfonts} 
\usepackage{amssymb}
\usepackage{latexsym}
\usepackage{hyperref}
\usepackage[english]{babel}
\usepackage[utf8]{inputenc}
\usepackage{xcolor}
\usepackage{feynmp}
\usepackage{bm}
\usepackage{bbold}
\usepackage{slashed}
\usepackage{bm}  % To make greek letters in bold face {\bm \Phi}
\usepackage{comment}
\usepackage{threeparttable}

\usepackage{setspace}
\allowdisplaybreaks

\usepackage{bigstrut}

\begin{document}

{\onehalfspacing
\begin{center}

{\Large\bf Limits on the Carroll-Field-Jackiw electrodynamics from geomagnetic data}

\vspace*{0.3cm}

G. F. de Carvalho$^{1}$\\
\vspace*{0.1cm}
{Instituto de F\'{\i}sica, Universidade Federal do Rio de Janeiro\\
 Av. Athos da Silveira Ramos, 149
Centro de Tecnologia - bloco A - Cidade Universitária - Rio de Janeiro - RJ - CEP: 21941-909}

\vspace*{0.3cm}

M. Fillion$^{2}$\\
\vspace*{0.1cm}
{Universit\'e Paris Cit\'e, Institut de physique du globe de Paris, CNRS, F-75005 Paris, France}\\

\vspace*{0.3cm}

P. C. Malta$^{3}$\\
\vspace*{0.1cm}
{R. Antonio Vieira 23, 22010-100, Rio de Janeiro, Brazil }\\

\vspace*{0.3cm}

C. A. D. Zarro$^{4}$\\
\vspace*{0.1cm}
{Instituto de F\'{\i}sica, Universidade Federal do Rio de Janeiro\\
 Av. Athos da Silveira Ramos, 149
Centro de Tecnologia - bloco A - Cidade Universitária - Rio de Janeiro - RJ - CEP: 21941-909}

\vspace*{0.2cm}
\end{center}

\begin{abstract}
Lorentz-symmetry violation may be described via the CPT-odd, dimension-3, Carroll-Field-Jackiw term, which couples the electromagnetic fields to a constant 4-vector $k_{\rm AF}$ selecting a preferred direction in spacetime. We solve the field equations using the Green's method for a static point-like magnetic dipole and find the $k_{\rm AF}$-dependent corrections to the standard dipolar magnetic field that strongly dominates the near-Earth magnetic field. Given the very good agreement between current models and ground- and satellite-based geomagnetic data, our strongest constraints on the components of $k_{\rm AF}$ in the Sun-centered frame read $|(k_{\rm AF})_Z| \lesssim 4 \times 10^{-25} \, {\rm GeV}$ for $|(k_{\rm AF})_X|, |(k_{\rm AF})_Y| \lesssim 10^{-24} \, {\rm GeV}$ at the two-sigma level. This represents an improvement of about four orders of magnitude over earlier bounds based on other geophysical phenomena.

\noindent
  
\end{abstract}}

\vfill
\noindent\underline{\hskip 140pt}\\[4pt]
{$^{1}$ E-mail address: guilhermefreiredc@gmail.com} \\
\noindent
{$^{2}$ E-mail address: fillion@ipgp.fr} \\
\noindent
{$^{3}$ E-mail address: pedrocmalta@gmail.com} \\
\noindent
{$^{4}$ E-mail address: carlos.zarro@if.ufrj.br}

%%%%%%%%%%%%%%%%%%%%%%%%%%%%%%%
\section{Introduction} \label{sec_intro}
\indent

Lorentz invariance is a cornerstone of quantum field theory and it has withstood every test so far~\cite{Mattingly_2005}. Nonetheless, string-based extensions of the Standard Model (SM) may induce Lorentz-symmetry violation (LSV) at very high energies~\cite{Kostelecky:1988zi} in all sectors of the SM; these effects are systematically catalogued in the Standard Model Extension (SME)~\cite{Colladay:1996iz, Colladay_1998}. Since the development of the SME, experimental tests of LSV are generally stated within its framework and experimental limits on its various sectors are collected by Kosteleck\'y and Russel in Ref.~\cite{Kosteleck_2011}.

Carroll, Field and Jackiw (CFJ) introduced a CPT-odd, spacetime constant 4-vector $k_{\rm AF}$ via the gauge-invariant term~\cite{Carroll:1989vb}
\begin{equation}\label{eq_CFJ}
\mathcal{L}_{\rm CFJ} = \frac{1}{2} \epsilon_{\mu\nu\alpha\beta} \left(k_{\rm AF}\right)^{\mu} \! A^{\nu}F^{\alpha\beta} \; , 
\end{equation}
where the background 4-vector has canonical dimension of ${\rm energy}^{1}$, $A^{\mu}=\left(A_0, \, {\bf A} \right)$ is the electromagnetic 4-potential and the field-strength tensor is $F^{\mu\nu} = \partial^\mu A^\nu - \partial^\nu A^\mu$, as usual. There are, however, serious theoretical issues related to the stability, unitarity and causality of the theory for a purely time-like background~\cite{Adam_2001, Casana_2008}. Hence, only a space-like background is physically acceptable.

In their seminal work~\cite{Carroll:1989vb}, Carroll, Field and Jackiw recognized that the LSV background causes the polarization of light to spontaneously rotate in vacuum and in the absence of external fields. It is a cumulative effect and light from farther sources should produce larger rotations; the non-observation of such signals in data from radio galaxies allowed them to constrain the spatial components of the background to $\lesssim 10^{-42} \, {\rm GeV}$~\cite{Carroll:1989vb, Goldhaber:1996fw}. Incidentally, they also used geomagnetic data to limit a time component to $\lesssim 6 \times 10^{-26}$~GeV~\cite{Carroll:1989vb}. Most notably, bounds extracted from precise measurements of the cosmic background radiation (CMB) pose even stronger limits~\cite{Kostelecky:2008be, Pogosian:2019jbt}. Moving away from cosmological distances and closer to Earth, the analysis of deviations of the Amp\`ere-Maxwell law in solar-wind (satellite) data within Earth's magnetosphere gives $\lesssim 10^{-26} \, {\rm GeV}$~\cite{Spallicci_2024}, whereas putative shifts in the Schumann resonance frequencies in Earth's ionosphere lead to $\lesssim 10^{-20} \, {\rm GeV}$~\cite{Mewes_2008}.

Astrophysical bounds profit from the long path lengths of light from quasars, pulsars or the CMB~\cite{Ferreira_2025,Caloni}. However, it is possible that LSV vectors vary over cosmological scales, introducing further complexity in the modeling of the astrophysical environment~\cite{Mewes_2008, Kost_2008_1, Kostelecky:2005ic, Bailey:2006fd}. Shorter scales generally offer reduced sensitivities, but this is partly compensated by better experimental control and more robust modelling. Local tests are thus preferable: limits from the non-observation of polarization rotation in cavity experiments reach $\lesssim 10^{-22} \, {\rm GeV}$, whereas limits from hydrogen spectroscopy attain $\lesssim 10^{-19} \, {\rm GeV}$~\cite{Gomes:2016wqj}, both for spatial components.

The CFJ term~\eqref{eq_CFJ} modifies Maxwell's equations due to the constant background coupling to the electromagnetic fields, implying that the fields produced by charge and current distributions will differ from those in Maxwell's theory in interesting ways. Simple field configurations in unbounded $(3+1)$ dimensions have been studied~\cite{Borges_2022, Casana_2008}, but also in waveguides~\cite{Gomes_2010} and in $(2+1)$ dimensions~\cite{Belich_2005}. In the presence of the CFJ term with a purely time-like background $k^0_{\rm AF}$ the electrostatic fields generated by a point electric charge are unaffected by LSV, whereas the magnetic field due to a steady localized current is strongly modified by the time-like CFJ background, exhibiting periodic spatial variation with a long wavelength of $\sim 1/k^0_{\rm AF}$. On the other hand, for a purely space-like component, static electric charges generate not only electric fields, but also magnetic fields; likewise, steady currents generate electric fields. Interestingly, electric dipoles also experience an spontaneous torque tilting the system towards ${\bf k}_{\rm AF}$~\cite{Borges_2022}.

To the best of our knowledge, the field configuration produced by a static magnetic dipole has not been studied in the context of the CFJ electrodynamics. Here we remedy this omission. As it turns out, the magnetic field is composed of three parts: the standard dipolar field from Maxwell's electrodynamics, a first-order correction proportional to the time component of the CFJ background and second-order terms involving only its spatial components. The first LSV correction was investigated by Carroll, Field and Jackiw~\cite{Carroll:1989vb} -- our focus will be on the second-order terms, as these can have potentially detectable effects on the geomagnetic field data. In fact, such data have been used to constrain models of physics beyond the Standard Model, including magnetic monopoles~\cite{monopole}, spin-dependent long-range forces~\cite{Lai, Hunter1, Hunter2}, axion-like particles~\cite{Arza_2022, Sulai_2023}, dark photons~\cite{Fedderke_2021, Fedderke_2021_2, Sulai_2023} and massive photons~\cite{Schroedinger_1, Schroedinger_2, Goldhaber_1968, Fischbach:1994ir}.

This paper is organized as follows: in Sec.~\ref{sec_CFJ} we discuss the basics of CFJ electrodynamics and in Sec.~\ref{sec_dipole} we obtain the magnetic field of a magnetic dipole. Next, in Sec.~\ref{sec_geomag} we present relevant facts about the geomagnetic field which will be useful in Secs.~\ref{sec_limits} and~\ref{sec_sensis}, where we extract limits and project long-term sensitivities on $k_{\rm AF}$. Finally, in Sec.~\ref{sec_conc} we present our conclusions. The metric is $\eta^{\mu\nu} = {\rm diag}(1,-1,-1,-1)$, the totally antisymmetric Levi-Civita symbol $\epsilon^{\mu\nu\lambda\kappa}$ is defined with $\epsilon^{0123} = +1$ and $\epsilon^{0ijk} = -\epsilon_{0ijk} = \epsilon^{ijk} = \epsilon_{ijk}$. Four-vectors are defined as $V^\mu = (V^0, {\bf V})$ with $V^0 = V_0$; spatial components are denoted by ${\bf V}^k = V_k$ when confusion with the 4-vector cannot occur. Data are presented in SI units, but the calculations are performed in natural units, $c = \hbar = \epsilon_0 = 1$. For future convenience, magnetic fields are measured in ${\rm nT} = 10^{-9} \, {\rm T} = 1.95 \times 10^{-25} \, {\rm GeV}^2$, Earth's mean radius is $R_\oplus = 6371.2 \, {\rm km} = 3.23 \times 10^{22} \, {\rm GeV}^{-1}$ and magnetic moments have unit $1 \, {\rm A \cdot m^2} = 3.19 \times 10^{25} \, {\rm GeV}^{-1}$.

%%%%%%%%%%%%%%%%%%%%%%
\section{The CFJ electrodynamics} \label{sec_CFJ}
\indent

Adding $\mathcal{L}_{\rm CFJ}$~\eqref{eq_CFJ} to Maxwell's Lagrangian we have 
\begin{equation}
\mathcal{L} = -\frac{1}{4}  F_{\mu\nu} F^{\mu\nu} + \frac{1}{2} \epsilon_{\mu\nu\alpha\beta} \left(k_{\rm AF}\right)^{\mu} \! A^{\nu}F^{\alpha\beta} -  J_\mu A^\mu \, ,
\end{equation}
where $J^\mu = \left( \rho, \, {\bf J} \right)$ represent conserved currents. Conveniently denoting the background by $k_{\rm AF} \equiv k$ and varying $\mathcal{L}$ with respect to $A^{\alpha}$, we obtain the inhomogeneous equations of motion
\begin{equation}\label{eq_motion_A} 
\partial_{\beta}F^{\beta\alpha} + \epsilon^{\mu\alpha\beta\lambda} k_{\mu} F_{\beta\lambda}  =  J^{\alpha} \, .
\end{equation} 
Using $F^{0i} = -{\bf E}_i$ and $F^{ij} = -\epsilon_{ijk}{\bf B}_k$ we find Gauss' and Amp\`ere-Maxwell's equations
\begin{subequations}
\begin{eqnarray}
{\bm \nabla}\cdot{\bf E} & = & \rho + 2{\bf k} \cdot{\bf B} \, , \label{eq_motion_1} \\
{\bm \nabla}\times{\bf B} -\partial_t{\bf E} & = &  {\bf J} + 2k_0{\bf B} -2{\bf k}\times{\bf E} \; , \label{eq_motion_2} 
\end{eqnarray}
\end{subequations}
while the Bianchi identities $\epsilon^{\mu\nu\sigma\kappa} \partial_{\mu} F_{\sigma\kappa} = 0$ give the homogeneous equations ${\bm \nabla} \times {\bf E} + \partial_t{\bf B} = 0$ and ${\bm \nabla}\cdot{\bf B} = 0$.

We are ultimately interested in the electromagnetic fields produced by a given source configuration. However, before moving on to this calculation in the next section, we must address an important issue. When we mention components of the background in LSV theories, we have to first specify the frame in which these are measured. The aforementioned theoretical issues regarding stability, unitarity and causality concern the components of $k$ as measured in an inertial reference frame. A convenient choice is the so-called Sun-centered frame (SCF)~\cite{Kosteleck_2011, Kosteleck__2002, Bluhm_2003}, which is approximately inertial in the time scales of most experiments. The connection between the components of $k$ in the SCF and in the laboratory frame with time coordinate $t$ and spatial coordinates $i=\{x,y,z \}$ is discussed in detail in App.~\ref{app_SCF}. Keeping in mind the requirement that $k^T \equiv 0$ in the SCF, we have
\begin{equation} \label{eq_components}
k^0 \approx - {\bm \beta} \cdot {\bf k}^{\rm SCF} \quad {\rm and} \quad {\bf k}^i \approx  (\mathcal{R} \, {\bf k}^{\rm SCF})^i \, ,
\end{equation}
where ${\bm \beta}$ is the velocity vector of the laboratory relative to the SCF and $\mathcal{R}$ is a rotation matrix; both are time dependent, thus introducing the time modulations typical of theories with LSV. The velocities involved are generally small ($|{\bm \beta}| \lesssim 10^{-4}$), implying that the time component is suppressed relative to the spatial component and we have the  hierarchy ${\bf k}^2 \gg k^0|{\bf k}| \gg (k^0)^2$. This allows us to safely neglect effects quadratic in the time component, but not in the spatial ones.

%%%%%%%%%%%%%%%%%%%%%%%%%%%%
\section{Field configuration of a point-like magnetic dipole} \label{sec_dipole}
\indent

As will be discussed in Sec.~\ref{sec_geomag}, despite of its complexity, the geomagnetic field near Earth's surface ($r = R_\oplus$) is dominated by a dipolar term. It is therefore important to determine the electromagnetic fields produced by a point-like, static magnetic dipole located at Earth's center in the presence of a non-zero CFJ background. As previously remarked, the electromagnetic fields and potentials generated by electric charges (static or in uniform motion), electric dipoles and Dirac strings were analysed in Refs.~\cite{Borges_2022, Casana_2008}. The fields of a magnetic dipole, however, have not been addressed so far -- here we close this gap.

We wish to determine the electromagnetic fields produced by a point-like magnetic dipole placed at Earth's center. Re-writing Eq.~\eqref{eq_motion_A} explicitly in terms of the 4-potential and imposing the Lorenz gauge $\partial_\mu A^\mu = 0$, we find
\begin{equation} \label{eq_wave_eq}
\left( \Box \eta^{\alpha\lambda} + 2\epsilon^{\mu\alpha\beta\lambda} k_{\mu}\partial_\beta \right) A_\lambda (x) = J^\alpha (x) \, .
\end{equation}
Next, going to momentum space with $p^\mu = (p^0, {\bf p})$, Eq.~\eqref{eq_wave_eq} becomes
\begin{equation} \label{eq_wave_eq_p}
D^{\alpha\lambda}(p) \tilde{A}_\lambda (p) = -\tilde{J}^\alpha (p) 
\end{equation}
with $D^{\alpha\lambda}(p) = p^2 \eta^{\alpha\lambda} + 2i\epsilon^{\mu\alpha\beta\lambda} k_{\mu} p_\beta$. In order to obtain the 4-potential, we must evaluate the inverse of this operator, which is given up to and including $\mathcal{O}(k^2)$ by
 \begin{eqnarray}\label{eq_D_inv}
D^{-1}_{\mu\nu}(p) & \approx & \frac{1}{p^4} \Bigg\{ \eta_{\mu\nu}\left[ \left( p^2 - 4k^2 \right) + \frac{4(k\cdot p)^2}{p^2} \right] +  4k^2 \frac{p_\mu p_\nu}{p^2}   \nonumber \\
& + & 2i\epsilon_{\mu\nu\rho\sigma} k^\rho p^\sigma   + 4k_\mu k_\nu   - \frac{4 k\cdot p}{p^2}\left( k_\mu p_\nu + k_\nu p_\mu \right)  \Bigg\} \, .
\end{eqnarray}
The exact result is given in Refs.~\cite{Casana_2008, Ba_ta_Scarpelli_2003}; see also Ref.~\cite{Borges_2022}.

The magnetization of a magnetic dipole ${\bm \mu}$ located at ${\bf x}_0$ is ${\bf M}({\bf x}) = {\bm \mu} \delta^3 ({\bf x} - {\bf x}_0)$, which induces a 3-current ${\bf J}({\bf x}) = {\bm \nabla} \times {\bf M}({\bf x})$~\cite{jackson_classical_1999}. Since $J_0 = 0$, Eq.~\eqref{eq_wave_eq_p} tells us that the vector potential at a point ${\bf x}$ is\footnote{In manifestly covariant notation the magnetization tensor is $M^{\mu\nu} = \epsilon^{\mu\nu\alpha\beta} u_\alpha M_\beta$ with $u^\alpha = (1,0)$ for a static dipole with magnetization $M^\beta = (0, {\bf M})$; the 4-current is then $J^\alpha = \partial_\beta M^{\beta\alpha}$~\cite{cross2012, jackson_classical_1999}.}
\begin{equation}
{\bf A}_i({\bf x}) = \int \! d^3 {\bf x}' {\bf J}_j ({\bf x}') \! \int \! \frac{d^3 {\bf p}}{(2\pi)^3} D^{-1}_{ij}({\bf p}) \, e^{i {\bf p}\cdot({\bf x} - {\bf x}')} \, . \label{eq_Ai}
\end{equation}
From Eq.~\eqref{eq_D_inv} with $k^\mu = (k^0, {\bf k})$ and assuming static sources $p^\mu = (0, {\bf p})$, we find~\cite{Casana_2008}
\begin{eqnarray}
D^{-1}_{ij}({\bf p}) & = &  \frac{1}{{\bf p}^2}\delta_{ij} - 2i\epsilon_{ijn} \frac{k_0 {\bf p}_n}{{\bf p}^4} - \frac{4}{{\bf p}^4} \Bigg\{ \delta_{ij} \left[ {\bf k}^2  \frac{({\bf k}\cdot{\bf p})^2}{{\bf p}^2}  \right]    \nonumber \\
& - & {\bf k}^2\frac{ {\bf p}_i {\bf p}_j }{{\bf p}^2} - {\bf k}_i {\bf k}_j + \frac{( {\bf k}\cdot{\bf p} )}{ {\bf p}^2 } \left( {\bf k}_i {\bf p}_j  + {\bf k}_j {\bf p}_i \right)   \Bigg\}   \, .  \label{eq_D_ij} 
\end{eqnarray}
A general expression for the vector potential is obtained in App.~\ref{app_details}, from which we find
\begin{eqnarray}
{\bf A}({\bf x}) & = & \frac{ {\bm \mu} \times \hat{{\bf r}} }{4\pi r^2}  + \frac{k_0}{4\pi r} \left[ {\bm \mu} + \left( {\bm \mu}\cdot\hat{{\bf r}} \right)\hat{{\bf r}}  \right] + \frac{3}{8\pi} \bigg\{ \left[ {\bf k}^2 + \frac{1}{3}\left(  {\bf k}\cdot\hat{{\bf r}} \right)^2  \right] \left( \hat{{\bf r}} \times {\bm \mu}  \right) \nonumber \\
& + & \left[ {\bf k} + \frac{1}{3}\left(  {\bf k}\cdot\hat{{\bf r}} \right)\hat{{\bf r}}  \right] \left[ {\bf k} \cdot \left( \hat{{\bf r}} \times {\bm \mu} \right) \right] + \left(  {\bf k}\cdot\hat{{\bf r}} \right) \left( {\bf k} \times {\bm \mu} \right) \bigg\} \, ,  \label{eq_Ai_final}
\end{eqnarray}
where ${\bf r} = {\bf x} - {\bf x}_0$, $r = |{\bf r}|$ and $\hat{{\bf r}} = {\bf r}/r$; the first term is the Lorentz-preserving result~\cite{jackson_classical_1999}. The scalar potential is finite and may be calculated in a similar fashion. The associated CFJ electric field is obtained in App.~\ref{app_E_field}.

The magnetic field is determined from the vector potential as usual via ${\bf B} = {\bm \nabla} \times {\bf A}$. As it turns out, it is composed of three pieces: the first is the Maxwellian dipolar field~\cite{jackson_classical_1999}
\begin{equation} \label{eq_B_field_Max}
{\bf B}_{\rm M}({\bf x}) = \frac{1}{4\pi r^3} \left[ 3\left(  {\bm \mu} \cdot\hat{{\bf r}} \right)\hat{{\bf r}} - {\bm \mu}  \right] \, ,
\end{equation}
which is corrected by the second and third pieces, much smaller, exclusively CFJ-originated contributions
\begin{eqnarray} \label{eq_B_field_CFJ}
{\bf B}_{\rm CFJ}({\bf x}) & = & \frac{k_0}{2\pi r^2} {\bm \mu} \times \hat{{\bf r}}  + \frac{1}{4\pi r} \bigg\{ \frac{3}{2} \left[ {\bf k}^2 - \left(  {\bf k}\cdot\hat{{\bf r}} \right)^2 \right] {\bm \mu}   \nonumber \\
& + & \left[  \left( {\bm \mu}\cdot{\bf k} \right) \left(  {\bf k}\cdot\hat{{\bf r}} \right) - \frac{3}{2} \left[ {\bf k}^2 + \left(  {\bf k}\cdot\hat{{\bf r}} \right)^2 \right] \left(  {\bm \mu}\cdot\hat{{\bf r}} \right)    \right] \hat{{\bf r}}   \nonumber \\
& + & \left[ \left(  {\bm \mu}\cdot\hat{{\bf r}} \right) \left(  {\bf k}\cdot\hat{{\bf r}} \right)  - 3 \left(  {\bm \mu}\cdot{\bf k} \right)\right] {\bf k} + 2 \left[ \hat{{\bf r}}\cdot \left(  {\bf k} \times {\bm \mu} \right) \right] \left( {\bf k} \times \hat{{\bf r}} \right) \bigg\}  \, .
\end{eqnarray}
The first term above, linear in $k_0$, was already studied by Carroll, Field and Jackiw in Ref.~\cite{Carroll:1989vb}, where the authors noted that it gives rise to an azimuthal magnetic field component. The second-order terms involving ${\bf k}$ offer a much richer phenomenology, since the axial symmetry around ${\bm \mu}$ present in ${\bf B}_{\rm M}$ is now broken. In fact, ${\bm \mu}$ and ${\bf k}$ select preferential directions and rotational symmetry is generally lost. Thus, besides the components along $\hat{{\bf r}}$ and ${\bm \mu}$, there are new terms $\sim {\bf k}$ and $\sim {\bf k} \times \hat{{\bf r}}$ pointing at unknown directions, thus allowing the CFJ part of the magnetic field to point in any direction at a given location, thereby acting to modify the effect of the usual dipolar field.

%%%%%%%%%%%%%%%%%%%%%%%%%%%%
\section{The geomagnetic field} \label{sec_geomag}
\indent

The near-Earth geomagnetic field orignates from electric currents and magnetized rocks in a variety of environments, from inside the solid Earth in the outer core, up to several Earth radii away from the surface deep in the magnetosphere. The field produced by the motion of electrically conducting fluid in the outer core is responsible for most of the field strength at the surface and is referred to as the ``main field"~\cite{Landeau_2022, Bruce_2000}. Its variation at time scales from months to millions of years remains difficult to predict~\cite{Fournier_2010}. Near the Earth, far from its sources, the main magnetic field can be expressed as ${\bf B} = - {\bm \nabla} V_{\rm int}$, where the scalar magnetic potential 
\begin{equation} \label{eq_V_int}
V_{\rm int} = R_\oplus \sum_{n=1}^{N} \sum_{m=0}^{n} \left( \frac{R_\oplus}{r} \right)^{n+1} P_n^{\, m} \left( \cos\theta \right) \big[ g_n^m (t) \cos\left( m\phi \right) + h_n^m (t) \sin\left( m\phi \right) \big]  \, 
\end{equation}
satisfies $\nabla^2 V_{\rm int} = 0$. The Gauss coefficients $\{g_n^m, h_n^m \}$ carry a linear time dependence to account for secular variation and $P_n^{\, m} \left( \cos\theta \right)$ are Schmidt semi-normalized associated Legendre functions of degree $n$ and order $m$~\cite{Winch}. For the main field usually $N \leq 13$ is enough to capture the features ($\approx 3000$~km) detected by satellites. We generally refer to potentials as $V_{\rm int}$ as internal magnetic potentials, and the corresponding magnetic fields as internal potential magnetic fields \cite{Sabaka_Hulot_Olsen_2010}.

Data from ground stations and satellite missions are used by different international collaborations to produce global empirical models of the main field, such as the World Magnetic Model 2025-2030 (WMM-2025)~\cite{WMM2025} and the International Geomagnetic Reference Field, 13th generation (IGRF-13)~\cite{IGRF}. From the multipole expansion of the internal field~\eqref{eq_V_int} it follows that $\approx 93\%$ of the main field is dipolar~\cite{Campbell, IGRF}. The magnitude of the dipole moment $\mu = |{\bm \mu}|$ can be calculated using the first Gauss coefficients as~\cite{IGRF, RefFrames}
\begin{equation} \label{eq_mu_gauss}
\mu = \left( \frac{4\pi}{\mu_0} \right) R_\oplus^3 \sqrt{ (g_1^0)^2 + (g_1^1)^2 + (h_1^1)^2 } \, .
\end{equation}
Using $g_1^0 = -29351.8$~nT, $g_1^1 = -1410.8$~nT and $h_1^1 = +4545.4$~nT for epoch 2025.0 from the WMM-2025~\cite{WMM2025}, we find $\mu \approx 7.7 \times 10^{22} \, {\rm A \cdot m^2}$ pointing from (geographic) North to South with a slight tilt relative to the rotational axis; this tilt will be henceforth ignored. The intensity of the dipole-like field on the surface is $\approx 20-70 \, {\rm \mu T}$~\cite{WMM2025} with secular variations of $\approx 80$~nT/year~\cite{IGRF}.

Beside the main field, one of the largest contributions to measurements, particularly near the surface, are lithospheric and crustal fields.~\cite{Thebault, ref_litho}. These are produced by magnetized rocks and carry information on the composition and geological history of the crust, being essentially frozen in the time scales relevant here~\cite{THOMPSON198461} and having spatial scales varying from a few kilometers to continental distances~\cite{map_crust}. This means that an expansion in spherical harmonics such as that of Eq.~\eqref{eq_V_int} must take into account much higher orders to deliver a faithful representation of all spatial scales that can be detected from extensive ground, marine, aeromagnetic and satellite data. For the World Digital Magnetic Anomaly Map (WDMAM) this means $N = 1050$, leading to a resolution as high as~40~km~\cite{WDMAM}. The crustal field is spatially highly inhomogeneous: from Table~1 of Ref.~\cite{WDMAM} we see that, while the average fields over the considered regions are relatively small, their spread can be as large as $\approx 100$~nT with maxima up to $\approx 4000$~nT. Ground magnetic measurements are thus particularly sensitive to crustal fields and can sense potentially strong small-scale fields from the underlying bedrock.

Finally, ionospheric and magnetospheric fields, generated by currents in the ionosphere and magnetosphere, are two other important contributions. These fields result from the interplay between the main field, the solar wind and the thermosphere-ionosphere environment, creating a complex time-dependent signature~\cite{Maus, Cowley, Yamazaki_2017, Richmond_1979, Millan_2017}. Near-Earth ionospheric and magnetospheric fields can also be modelled using spherical-harmonics-based approaches requiring more sophisticated parametrizations to account for their complex time variations and the complex morphology of ionospheric fields. Additionally, such models must also account for secondary fields generated by electric currents induced in the crust and mantle by time variations of ionospheric and magnetospheric fields~\cite{Sabaka_2020, olsem, Finlay_2020}. Ionospheric fields are weakest during local nighttime, when the ionospheric electrical conductivity is reduced, and display sidereal and annual time modulations. At the surface during nighttime, horizontal ionospheric fields have intensities $\lesssim 5$~nT~\cite{olsem}, whereas magnetospheric fields along the North-South direction are $\lesssim 100$~nT~\cite{Fischbach:1994ir, LANGEL1996235, olsem}. Geomagnetic models of the main field try to exclude crustal, ionospheric, magnetospheric and induced fields by adopting data selection and pre-processing strategies to minimize their effects, or by including contributions of these fields in the models to improve their separation~\cite{WMM2025}.

%%%%%%%%%%%%%%%%%%%%%%
\section{Limits on the CFJ background} \label{sec_limits}
\indent

Our main result is the CFJ magnetic field, ${\bf B}_{\rm CFJ}$, cf. Eq.~\eqref{eq_B_field_CFJ}, which is expected to modify the geomagnetic field as measured on or near Earth's surface. Importantly, as discussed in App.~\ref{app_SCF}, the CFJ background changes over time, thereby endowing ${\bf B}_{\rm CFJ}$ with a time dependence that is sensitive to the path of the measurement apparatus relative to the SCF. A major consequence of this is that  
observations made on ground-based or satellite-borne platforms will in principle be able to access different combinations of the CFJ background~\cite{Bluhm_2003}. For these reasons, in the next sections we work out the explicit expressions of the CFJ magnetic field as ``seen" by an instrument fixed at a given location on the surface or onboard of a satellite to extract upper limits on the components of the CFJ background in the SCF.

%%%%%%%%%%%%%%%%%%%%%%
\subsection{Ground data} \label{sec_limit_ground}
\indent

For ground stations we work with a standard reference frame fixed to Earth's surface with spatial coordinates $\{x,y,z\}$ where the $x$-axis points South ($\hat{{\bf x}} = \hat{{\bm \theta}}$), the $y$-axis eastward ($\hat{{\bf y}} = \hat{{\bm \varphi}}$), and the $z$-axis vertically upward ($\hat{{\bf z}} = \hat{{\bf r}}$), cf. App.~\ref{app_lab}. Expressed in spherical coordinates, Earth's magnetic dipole moment is 
\begin{equation} \label{eq_def_mu}
{\bm \mu}/\mu = -\cos\theta \hat{\bf r} + \sin\theta \hat{\bm \theta} \, ,
\end{equation}
where the co-latitude $\theta$ is the usual polar angle measured from the geographic North Pole and $\mu$ is given by Eq.~\eqref{eq_mu_gauss}. The components of ${\bf B}_{\rm CFJ}$, cf. Eq.~\eqref{eq_B_field_CFJ}, read
\begin{eqnarray}
{\rm B}_{\rm CFJ, r} & = & \frac{\mu k_z}{2\pi r} \left( 2k_z c_\theta - k_x s_\theta \right) \, , \label{eq_B_radial_lab} \\
{\rm B}_{\rm CFJ, \theta} & = & \frac{\mu}{8\pi r} \left[ 4k_z k_x c_\theta - \left(3k_x^2 + k_y^2 \right) s_\theta \right]  \, , \label{eq_B_theta_lab} \\
{\rm B}_{\rm CFJ, \varphi} & = & -\frac{\mu k_0 s_\theta}{2\pi r^2} + \frac{\mu k_y}{4\pi r} \left( 2k_z c_\theta - k_x s_\theta \right) \label{eq_B_phi_lab}
\end{eqnarray}
with $\{k_0, k_x, k_y, k_z \}$ described in terms of the SCF components $\{ k_X, k_Y, k_Z \}$ by Eqs.~\eqref{eq_app_kx_lab}-\eqref{eq_app_kz_lab}. Henceforth we adopt the notation $s_\gamma \equiv \sin\gamma$ and $c_\gamma \equiv \cos\gamma$ when convenient.

Finally, it is instructive to estimate the order of magnitude of the attainable bounds. The second-order terms have the form $\sim \mu {\bf k}^2/4\pi r$ and will be constrained to be smaller than $\delta\mathcal{B}$, typically a difference between observation and a Lorentz-symmetry preserving model. We may thus write
\begin{equation} \label{eq_estimate}
|{\bf k}| \lesssim 6 \times 10^{-25} \, {\rm GeV} \sqrt{ \left( \frac{ r }{ R_\oplus } \right) \left( \frac{ \delta\mathcal{B}}{10 \, {\rm nT}} \right) } \, ,
\end{equation}
which, assuming the $\delta\mathcal{B} \approx 100$~nT typical of ground data, gives $|{\bf k}| \lesssim 2 \times 10^{-24} \, {\rm GeV}$. As we will see, this is at the ballpark of most of our bounds.

%%%%%%%%%%%%%%%%%%%%%%
\subsubsection{Azimuthal fields} \label{sec_limit_phi_ground}
\indent

As briefly mentioned at the end of Sec.~\ref{sec_dipole}, the first term in Eq.~\eqref{eq_B_field_CFJ}, the one involving the time component of the CFJ background, $k_0$, would induce an azimuthal component to the main dipolar field, which originally does not contain such an East-West component. Carroll, Field and Jackiw studied this contribution in their seminal paper and noted that this azimuthal component was not observed. More precisely, they mention that time-varying ionospheric and/or magnetospheric fields do not allow the internal field to be determined with precision better than $\approx 10$~nT. Thus, assuming that the standard Lorentz-preserving theory matches well the observations, one may impose that any LSV effect must be smaller than $\approx 10$~nT, leading to $k_0 \lesssim 6 \times 10^{-26}$~GeV~\cite{Carroll:1989vb}.

In their work, which predates the SME by seven years, the time-modulation of the background 4-vector as seen from an Earth-bound laboratory was not considered. Actually, as shown in App.~\ref{app_lab}, $k_0$ is affected by the orbital velocity of the laboratory relative to the SCF, modulated by a frequency $\Omega_\oplus = 2\pi/{\rm year}$, cf. Eq.~\eqref{eq_app_k0_lab}. Therefore, from Eq.~\eqref{eq_B_phi_lab} we see that, over several years, the term proportional to $k_0$ averages to zero. However, instead of taking time averages, they considered the largest possible LSV effect and constrained it to be smaller than~10~nT; incidentally, this still remains a reasonable estimate~\cite{Maus}. Assuming that $s_\Psi, c_\Psi \sim \mathcal{O}(10^{-1})$ with $\Psi = \Omega_\oplus T$, their limit translates into 
\begin{equation} \label{eq_limit_k0_ground_CFJ}
|k_X -  0.9 k_Y - 0.4 k_Z | \lesssim 6 \times 10^{-21} \, {\rm GeV} \, .
\end{equation}

\begin{figure}[t!]
\centering
\begin{minipage}[b]{0.80\linewidth}
\includegraphics[width=\textwidth]{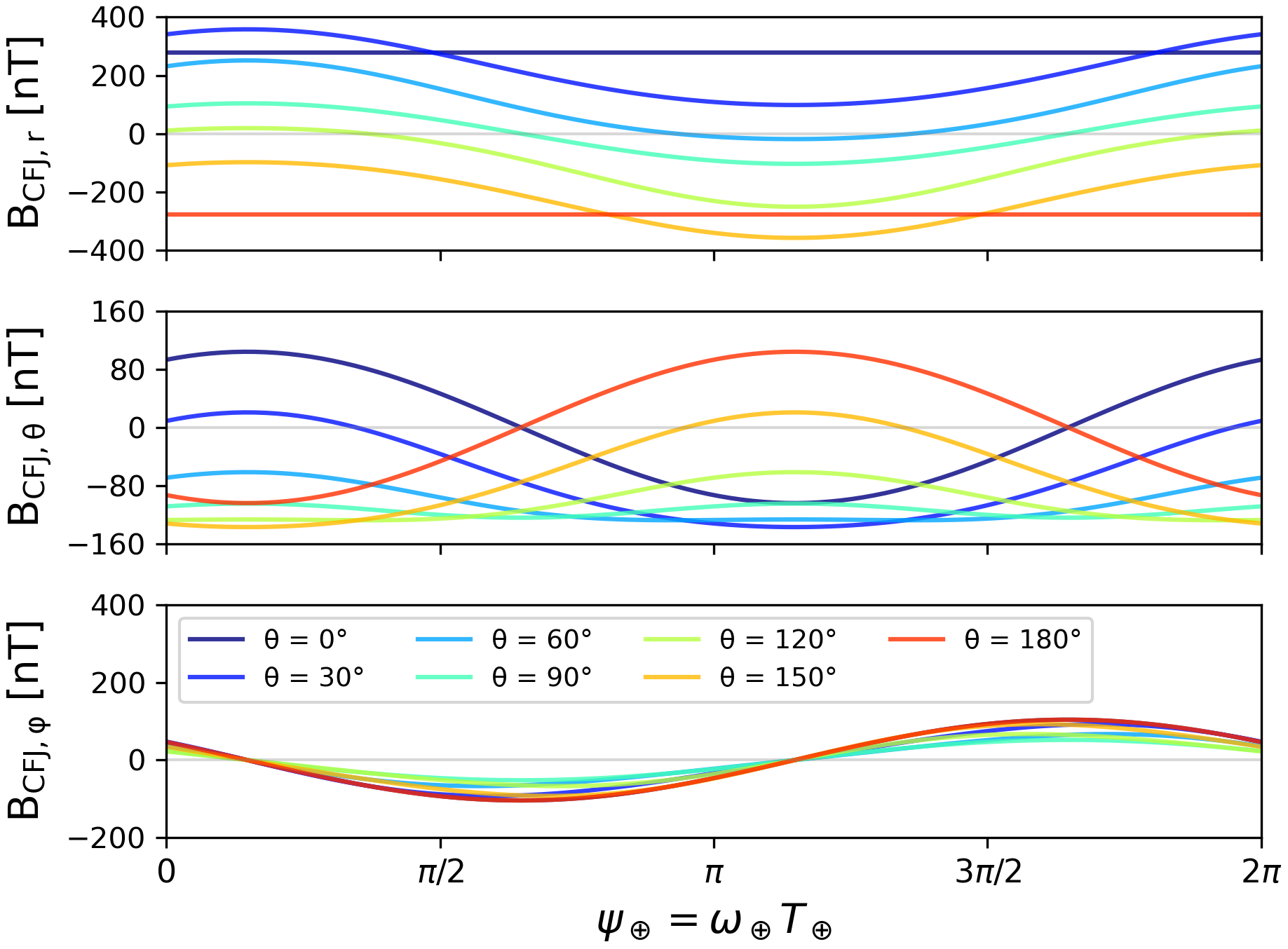}
\end{minipage} \hfill
\caption{From top to bottom: time variation of the radial, polar and azimuthal (only its quadratic part) components of the CFJ field, cf. Eqs.~\eqref{eq_second_term_phi}, \eqref{eq_B_r_lab_full} and~\eqref{eq_B_theta_lab_full}, over a sidereal day for different co-latitudes. Here we set $k_X = 1$, $k_Y = 0.5$ and $k_Z = 1.5$ in units of $10^{-24} \, {\rm GeV}$.}
\label{fig_B_time}
\end{figure}

Regarding the second-order corrections, since ${\bf k}$ has an unknown orientation, the terms proportional to ${\bf k}$ and ${\bf k} \times \hat{{\bf r}}$ will generally also have azimuthal components. The combination of these components yields 
\begin{equation}  \label{eq_second_term_phi}
{\rm B}_{\rm CFJ, \varphi}^{k^2} = \frac{\mu }{8\pi R_\oplus} \Bigg\{ s_{2\theta} \left[ c_{2\psi_\oplus} k_X k_Y  - \frac{s_{2\psi_\oplus}}{2}\left( k_X^2 - k_Y^2 \right)  \right] - \left( 3 + c_{2\theta} \right) \left( s_{\psi_\oplus}k_X k_Z  - c_{\psi_\oplus}k_Y k_Z  \right) \Bigg\} 
\end{equation}
and is shown in the lower panel of Fig.~\ref{fig_B_time}. Time is encoded in $\psi_\oplus = \omega_\oplus T_\oplus$ with $\omega_\oplus = 2\pi/{\rm day}$, showing that this field varies daily. Note that $T_\oplus = T - T_{\oplus,0}$, where $T_{\oplus,0}$ is an initial time that is specific to each experiment, cf. App.~\ref{app_lab}. Most importantly, ${\rm B}_{\rm CFJ, \varphi}^{k^2}$ averages to zero over periods of many days. In conclusion, the azimuthal CFJ field as measured by an Earth-bound observer would contribute at most to the daily or yearly fluctuations of the overall magnetic field with no finite offset in the long run.

%%%%%%%%%%%%%%%%%%%%%%
\subsubsection{Radial and polar fields} \label{sec_limit_radial_polar_ground}
\indent

\begin{figure}[t!]
\centering
\begin{minipage}[b]{0.80\linewidth}
\includegraphics[width=\textwidth]{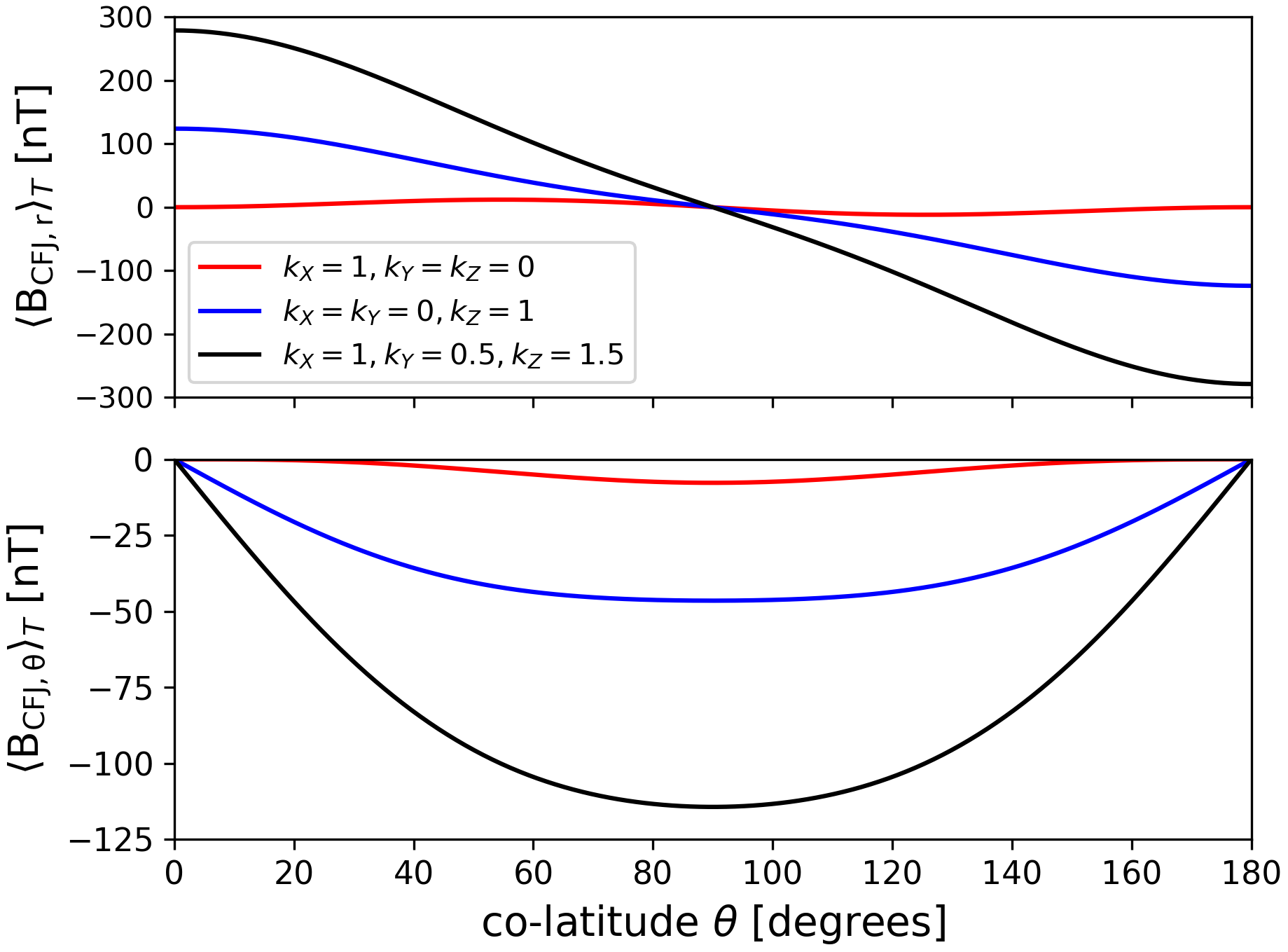}
\end{minipage} \hfill
\caption{Time-averaged radial and polar components of the CFJ field for an Earth-bound observer at the surface at $r = R_\oplus$: radial~\eqref{eq_Br_avg_T} in the top panel and polar~\eqref{eq_Btheta_avg_T} in the bottom panel. The components of the background in the SCF are expressed in units of $10^{-24} \, {\rm GeV}$. Since the polar angle satisfies $0 \leq \theta \leq \pi$, the radial projection will change sign when crossing the geographic equator due to the global factor of $c_\theta$, whereas the polar projection is negative for any co-latitude. }
\label{fig_map}
\end{figure}

Now we turn to the more interesting case of the radial and polar components as given by Eqs.~\eqref{eq_B_radial_lab} and~\eqref{eq_B_theta_lab}. Plugging in the expressions for the components of the CFJ background in the standard Earth-bound laboratory frame in terms of those in the SCF, cf. Eqs.~\eqref{eq_app_kx_lab}-\eqref{eq_app_kz_lab}, we find
\begin{eqnarray}
{\rm B}_{\rm CFJ, r} & = & \frac{\mu}{4\pi r} \Big[ 2c_\theta s^2_\theta (c_{\psi_\oplus}^2 k_X^2 + s_{\psi_\oplus}^2 k_Y^2) + \frac{1}{2} (7c_\theta + c_{3\theta})k_Z^2  \nonumber \\
& + & 2c_\theta s^2_\theta s_{2\psi_\oplus}k_X k_Y + (3s_\theta + s_{3\theta}) (c_{\psi_\oplus} k_X k_Z + s_{\psi_\oplus} k_Y k_Z)  \Big] \, , \label{eq_B_r_lab_full} \\
{\rm B}_{\rm CFJ, \theta} & = & \frac{\mu}{32\pi r} \Big[ s_\theta (-1 + 2c_{2\theta}c_{\psi_\oplus}^2 + 3c_{2\psi_\oplus}) k_X^2 - s_\theta (1 - 2c_{2\theta}s_{\psi_\oplus}^2 + 3c_{2\psi_\oplus}) k_Y^2 - (13s_\theta + s_{3\theta}) k_Z^2 \nonumber \\
& + & s_{2\psi_\oplus} (5s_\theta + s_{3\theta}) k_X k_Y + 2 (7c_\theta + c_{3\theta}) (c_{\psi_\oplus}k_X k_Z + s_{\psi_\oplus}k_Y k_Z) \Big] \, ,  \label{eq_B_theta_lab_full}
\end{eqnarray}
which are shown in the top and center panels of Fig.~\ref{fig_B_time}, respectively. The relatively short period of the LSV signal and the much longer time span of the available magnetic data allow us to consider time averages $\langle \dots \rangle_{T}$, resulting in
\begin{eqnarray} 
\langle {\rm B}_{\rm CFJ, r} \rangle_{T} & = & \frac{\mu c_\theta}{4\pi r} \left[ s^2_\theta \left(k_X^2 + k_Y^2 \right) + \left( 3 + c_{2\theta} \right)k_Z^2 \right] \, ,  \label{eq_Br_avg_T} \\
\langle {\rm B}_{\rm CFJ, \theta} \rangle_{T} & = & -\frac{\mu s_\theta}{16\pi r}  \left[ s_\theta^2 \left( k_X^2 + k_Y^2 \right) + \left( 7 + c_{2\theta} \right)k_Z^2 \right] \, .  \label{eq_Btheta_avg_T} 
\end{eqnarray}
Contrary to the azimuthal components, here we do have finite offsets. These results are shown in Fig.~\ref{fig_map} for $r = R_\oplus$.

After averaging over a number of sidereal days, it is clear that $k_X$ and $k_Y$ appear in the particular combination $k_X^2 + k_Y^2$, neatly separated from $k_Z^2$. This can be explained by the orientation of the axes of the SCF relative to Earth's equatorial plane, cf.~App.~\ref{app_SCF}. Since the $X$- and $Y$-axes of the SCF are parallel to the equator and Earth's spin remains aligned with the $Z$-axis, we have a degeneracy in the $X-Y$ plane preventing us from separately constraining $k_X$ or $k_Y$. In the next section we will see how our analysis can benefit from considering data from dedicated satellites in low-Earth orbits, allowing us to access different combinations of $\{ k_X, k_Y, k_Z \}$.

A question now arises: is it possible to absorb the particular LSV signatures from the CFJ electrodynamics into the multipole expansions of the geomagnetic field? If so, the resulting Gauss coefficients would include information on the CFJ background and we would be unable to extract bounds by comparing the residual difference between measurements and the modelled field. From a purely theoretical standpoint,  the answer is negative. The main reason is that the modified Amp\`ere-Maxwell equation~\eqref{eq_motion_2} gives ${\bm \nabla}\times{\bf B}_{\rm CFJ} \neq 0$, implying that the CFJ magnetic field is not a potential field and cannot be represented as in Eq.~\eqref{eq_V_int} in terms of a scalar magnetic potential satisfying Laplace's equation. Nevertheless, the situation is not as straightforward and further practical considerations must be taken into account.

The key observation lies in the dependence of potential and CFJ fields with $r$. The CFJ field decays as $\sim 1/r$, whereas internal fields decay faster as $\sim 1/r^{n+2}$ with $n \geq 1$~\cite{WMM2025, Fillion_2}. Incidentally, ionospheric or magnetospheric fields behave differently and are generally represented as external magnetic fields, which vary as $\sim r^{n-1}$~\cite{Sabaka_Hulot_Olsen_2010}. 
A problem arises if all measurements used to determine the Gauss coefficients of the scalar magnetic potentials were performed at exactly the same radius. In such a configuration, we would most likely face a uniqueness issue, as it would be very difficult to unambiguously  separate mathematically the CFJ terms from a degenerated  multipole expansion of the geomagnetic field on a surface at a fix radius.

In models such as the IGRF-13 and WMM-2025, however, data from ground stations are complemented by satellite data~\cite{IGRF, Alken_IGRF13}. Ground data are acquired at $r \approx R_\oplus$ with variations of $\approx 10$~km, while satellite data are recorded higher and span a range of several hundred km at thermospheric altitudes, typically between 300 and 800 km.  Take, for example, the three Swarm satellites of the European Space Agency with initial mean altitudes of $\approx 450$~km (Swarm A and C) and $\approx 530$~km (Swarm B)~\cite{swarm}, and the CryoSat-2 satellite \cite{Olsen_Albini_Bouffard_Parrinello_Toffner-Clausen_2020} higher up at a mean altitude of about 700 km. Empirical models constructed with a combination of ground and satellite data spanning different altitudes have a better chance of rejecting CFJ-like terms in the residuals. Nevertheless, such a separation would still be challenging if variations of the CFJ term with altitude were too weak to be detected in the data. The CFJ term could, in this case, be mistaken for a weak degree-1 external field. We argue, however, that a CFJ field weak enough such that its altitude variations are below data error level would still be well constrained by the upper bound computed with our approach.

After clearing up the concerns above, let us return to the problem of finding upper bounds on the CFJ background using ground geomagnetic data. A first possibility is to use Ref.~\cite{Beggan}, where the author analyses annual means of ground data (1980-2021) to determine how a measurement on the surface would compare to the predictions of the IGRF-13. Defining the residuals as the difference between the measured fields and those predicted by the IGRF-13 model, the author finds that the means of the residuals are non-zero, in particular those of the polar and radial components, mainly due to the magnetospheric ring current, which is not modelled by the IGRF, but is nonetheless present in the means of the data. Moreover, the lack of an obvious variation with latitude in the residuals suggests that their spread (between $\approx 130-300$~nT) is due to crustal fields.

\begin{figure}[t!]
\centering
\begin{minipage}[b]{0.90\linewidth}
\includegraphics[width=\textwidth]{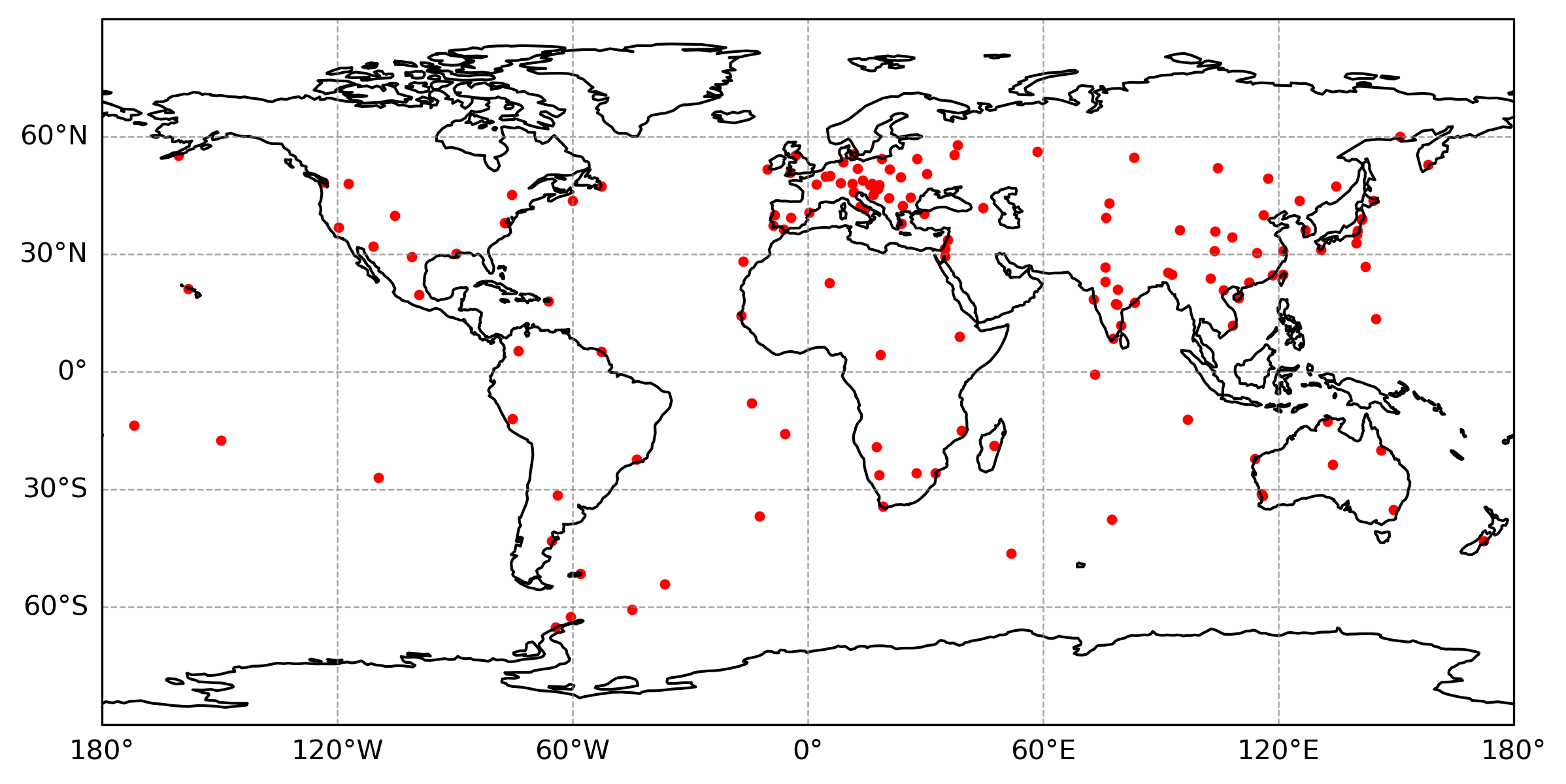}
\end{minipage} \hfill
\caption{Locations of the 144 ground observatories contributing to the International Real-Time Magnetic Observatory Network (INTERMAGNET) used in this section. }
\label{fig_map_obs}
\end{figure}

Our objective is to constrain the CFJ field, an assumedly very weak field. The aforementioned comparison, though adequate for the author's purposes, is not for ours. The reason is simple: while the purpose of Ref.~\cite{Beggan} was to quantify average errors at the Earth's surface with respect to the IGRF-13 and, therefore, construct a dataset representative of these errors, ours is to define a dataset that minimizes contributions from the same errors to unravel potential signals of the CFJ field. The most important contributions to the aforementioned errors are the ionospheric and magnetospheric fields, as well as higher-order lithospheric and crustal fields~\cite{IGRF}, effectively ignored by the IGRF-13 which is focused on the main geomagnetic field. However, these fields may be more accurately described by other modern models and accounting for them will be beneficial in our pursuit of strong bounds on the CFJ background.

To this end, we use hourly vector magnetic data from 144 ground observatories, shown in Fig.~\ref{fig_map_obs}, contributing to the International Real-Time Magnetic Observatory Network (INTERMAGNET)~\cite{intermag} and compiled by the British Geological Survey (BGS)~\cite{BGS} in the time period between January 1997 and December 2024. In total we have $\approx 1.31$ million vector data. A few selection criteria are imposed. Firstly, vector data are typically only collected for observatories in ``non-polar" latitudes, defined by quasi-dipole latitudes\footnote{Quasi-dipole coordinates trace along magnetic field lines~\cite{Richmond_1995}, but throughout this work we focus on standard geographic co-latitude, as is conventional in LSV studies. Nonetheless, the cut $\lambda_{\rm QD} < |55^\circ|$ is sensible to avoid auroral regions with much stronger magnetic activity that are usually worse modelled when compared to other regions at lower quasi-dipole latitudes.} $\lambda_{\rm QD} < |55^\circ|$; correspondingly, the lowest and highest geographic co-latitudes used here are $30^\circ$ and $155^\circ$, respectively. Secondly, in order to further limit perturbations from ionospheric electric currents, observatory data are taken exclusively during night time between 1 and 5 hours local time. Finally, data are only accepted at times when the Hp30 index~\cite{Hp30} is below one, thereby avoiding times of greater geomagnetic disturbance.

In order to remove remaining contributions of the core, lithospheric, and magnetospheric fields in the data, we use the version 8.3 of the CHAOS model series \cite{Kloss_Finlay_Olsen_Toffner-Clausen_Gillet_Grayver_2026}. It combines 25 years of geomagnetic data collected from satellites and ground stations and is used to produce the DTU candidate model for the IGRF-14 for epoch 2025. Covered in the model are the contributions from the core, lithospheric, and magnetospheric fields, as well as from secondary magnetic fields generated by electric currents induced by time variations of magnetospheric fields in the mantle. Additionally, we use the DIFI model~\cite{DIFI} to remove contributions from currents circulating in the E-region of the ionosphere, as well as secondary induced ionospheric fields. The latter step ensures that any remaining large-scale induced ionospheric fields are properly removed from the data, as this contribution was only partially removed when selecting data for nighttime. It is important to highlight that all the aforementioned model components are defined as potential fields represented as multipole expansions (Eq.~\eqref{eq_V_int}), thereby avoiding unintentional filtering of the LSV signal.

We constrain the CFJ background by comparing the LSV-induced magnetic field to the differences -- the residuals -- of data collected at ground observatories and the CHAOS-8 and DIFI-models~\cite{chaos_8, DIFI}, cf. Sec.~\ref{sec_geomag}. Given the amount of data and the typical time scale associated with the CFJ field, the background can be constrained by demanding that (with $u = \{r, \theta, \varphi\}$)
\begin{equation}  \label{eq_mean_std_def}
\langle {\rm B}_{\rm CFJ, u} \rangle = \langle \delta B_{u} \rangle \pm \sigma_{ \delta B_u } \, .
\end{equation}
Here $\delta B_{u}$ are the residuals and $\langle \dots \rangle$ denotes time and angle averaging, $\langle \delta B_{u} \rangle$ and $\sigma_{ \delta B_u }$ are the means and standard deviations. The average over co-latitude is not trivial due to the heterogeneous distribution of ground measurement stations, which are more concentrated in Europe and Asia, while the oceans and large portions of South America and Africa are more sparsely covered, cf. Fig.~\ref{fig_map_obs}.

Before discussing how we calculate the averages we must eliminate large outliers, potentially corresponding to measurements with especially large crustal biases. To this end, we use the interquartile range (IQR) as a measure of the spread of data. More concretely, for the original residue dataset we calculate the IQR, which is the difference of the $25\% \, (Q_1)$ and $75\% \, (Q_3)$ quantiles, that is, $IQR = Q_3 - Q_1$. Data are flagged as outliers and removed if they lie below $Q_1 - 3 \cdot IQR$ or above $Q_3 + 3 \cdot IQR$: from the initial $\approx 1.3$~million individual residuals, each with three components, $\approx 4\%$ were removed -- if the original data were normally distributed $\approx 0.02\%$ would have been excluded.

\begin{figure}[t!]
\centering
\begin{minipage}[b]{1.\linewidth}
\includegraphics[width=\textwidth]{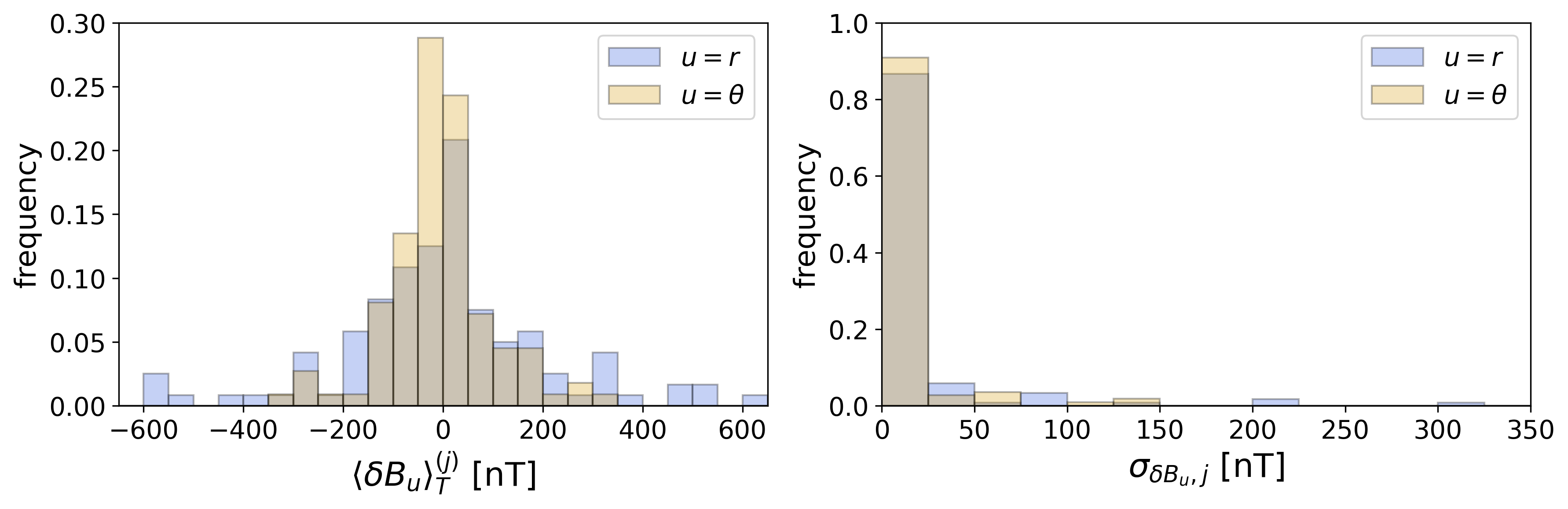}
\end{minipage} \hfill
\caption{Distributions of the filtered residuals in the non-empty angular patches of size $\Delta = 5^\circ$ following Eqs.~\eqref{eq_def_mean_residuals_T_obs} and~\eqref{eq_def_std_residuals_j_obs}. The height of each bin is the number of data in that bin divided by the total number; the sum of all heights is equal to one.  }
\label{fig_hist_obs}
\end{figure}

The quantities entering Eq.~\eqref{eq_mean_std_def} are defined as follows. Let us index individual filtered residuals by $i$ and divide the globe in square angular patches with sides of size $\Delta$ indexed by $j$; the area of each patch is $A_j = \int_{\rm patch \, j} \sin\theta d\theta d\phi$. The time and angle average of the filtered residuals is
\begin{equation} \label{eq_def_mean_residuals_obs}
\langle  \delta B_{u} \rangle = \displaystyle\sum_{j}  w_j \, \langle \delta B_{u} \rangle_{T}^{(j)}  \, ,
\end{equation}
where the normalized weights are $w_j = A_j/\sum_j A_j$; only non-empty patches are considered. The time average of the the filtered residuals $\langle \delta B_{u} \rangle_{T}^{(j)}$ within patch $j$ is
\begin{equation} \label{eq_def_mean_residuals_T_obs}
\langle \delta B_{u} \rangle_{T}^{(j)} = N_j^{-1} \displaystyle\sum_{i \, {\rm in \, patch \ j}} \, \delta B_{u, i} \, 
\end{equation}
with $N_j$ being the number of measurements in that patch -- if it is zero, this patch is not counted. Within each patch $j$ the (squared) standard deviation is 
\begin{equation} \label{eq_def_std_residuals_j_obs}
\sigma^2_{\delta B_{u},j} = (N_j - 1)^{-1} \displaystyle\sum_{i \, {\rm in \, patch \ j}} \left[ \delta B_{u, i} - \langle \delta B_{u} \rangle_{T}^{(j)} \right]^2  \, ,
\end{equation}
which can then be summed over all patches in quadrature to give 
\begin{equation} \label{eq_def_std_residuals_obs}
\sigma^2_{\delta B_{u}} = \sum_j w_j \sigma^2_{\delta B_{u},j} \, .
\end{equation}
The resulting weighted means and standard deviations of the filtered residuals of the radial and polar components are listed in Table~\ref{table_residuals_Fillion}; in Fig.~\ref{fig_hist_obs} we show the distributions of the residuals.

\begin{table}[t!]
\centering
\begin{tabular}{|c|c|c|c|c|}
\hline
$\langle  \delta B_{r} \rangle$ & $\sigma_{\delta B_r}$ & $\langle  \delta B_{\theta} \rangle$ & $\sigma_{\delta B_\theta}$  \\ \hline\hline
2.8 & 46.9 & -5.4 & 24.9  \\ \hline
\end{tabular}
\caption{Mean ($\langle  \delta B_{u} \rangle$) and standard deviation ($\sigma_{\delta B_u}$), with $u = \{ r, \theta \}$, of hourly residuals between ground data from INTERMAGNET observatories for the years 1997-2024 and the CHAOS-8 model. The data, after filtering with the IQR method, were weighted according to Eqs.~\eqref{eq_def_mean_residuals_obs} and~\eqref{eq_def_std_residuals_obs} using angular patches with sides of size $\Delta = 5^\circ$. All values are given in nT.}\label{table_residuals_Fillion}
\end{table}

\begin{figure}[t!]
\centering
\begin{minipage}[b]{0.80\linewidth}
\includegraphics[width=\textwidth]{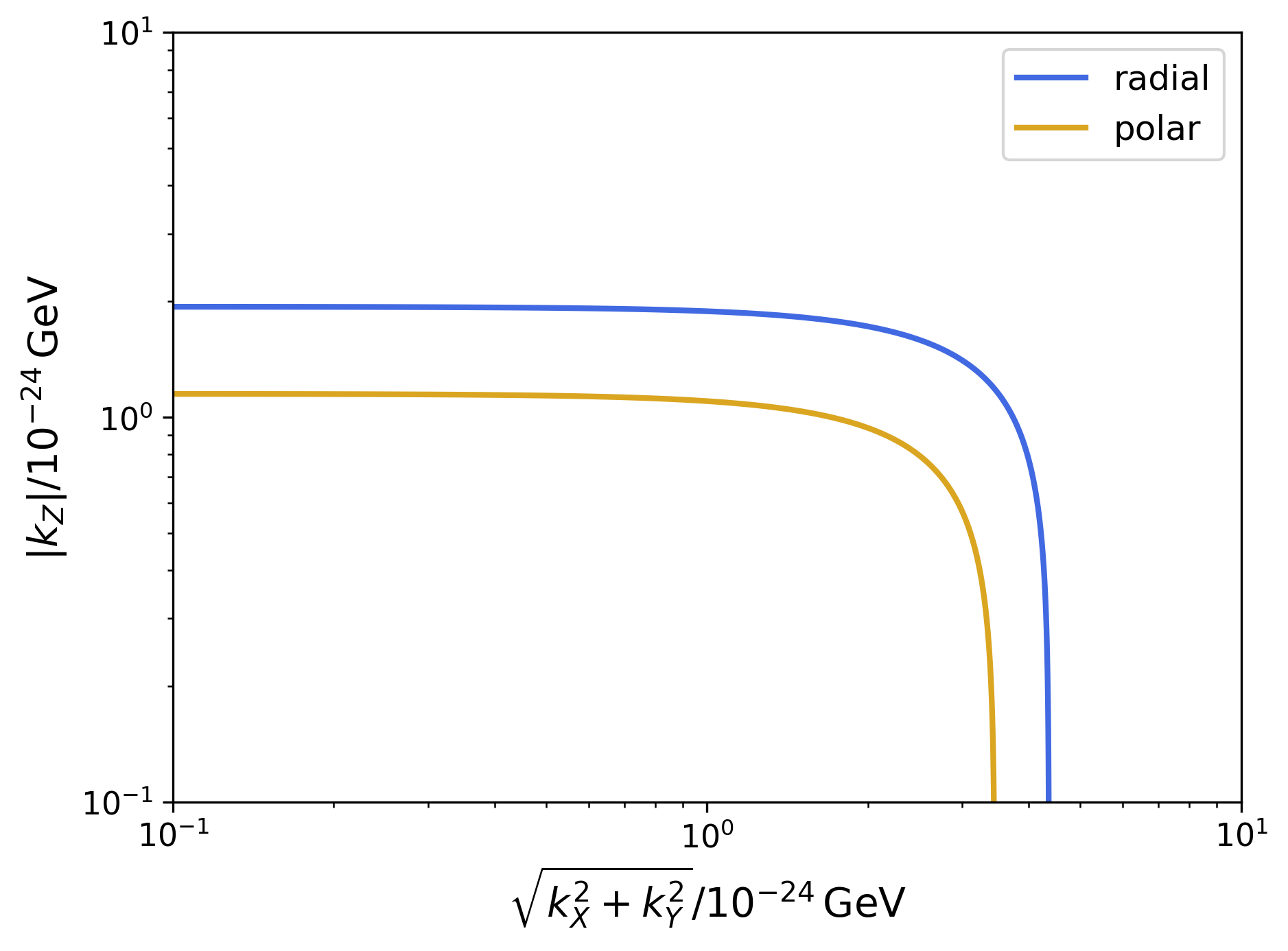}
\end{minipage} \hfill
\caption{Two-sigma bounds on combinations of the CFJ background in the SCF derived from ground data. In blue and yellow are the bounds from the radial and polar components, respectively. The region within the curves is allowed. }
\label{fig_bounds_joint}
\end{figure}

The left-hand side of Eq.~\eqref{eq_mean_std_def} can be evaluated using Eqs.~\eqref{eq_Br_avg_T} and~\eqref{eq_Btheta_avg_T} as
\begin{equation} \label{eq_def_mean_cfj_obs}
\langle {\rm B}_{\rm CFJ, u} \rangle = \displaystyle\sum_{j}  w_j \, N_{\rm obs}^{-1} \displaystyle\sum_{{\rm obs} \, {\rm in \, patch \, j}} \langle  {\rm B}_{\rm CFJ, u} \rangle_{T}(r_{\rm obs}, \theta_{\rm obs})  \, ,
\end{equation}
where $N_{\rm obs}$ is the number of observatories in patch $j$. Here we use the analytical time averages $\langle {\rm B}_{\rm CFJ, u} \rangle_{T}$. This is justified by our use of more than 25 years of hourly data and the fact that the CFJ field varies daily. Letting $k \equiv \tilde{k} \times 10^{-24} \, {\rm GeV}$ and $\Delta = 5^\circ$, the weighted averages become
\begin{eqnarray}
\langle {\rm B}_{\rm CFJ, r} \rangle & = & 5.1 \, {\rm nT} \left( \tilde{k}_X^2 + \tilde{k}_Y^2 + 5.0 \, \tilde{k}_Z^2  \right) \, , \label{eq_fillion_mean_r} \\
\langle {\rm B}_{\rm CFJ, \theta} \rangle & = & -4.6 \, {\rm nT} \left( \tilde{k}_X^2 + \tilde{k}_Y^2 + 9.0 \, \tilde{k}_Z^2  \right) \, .\label{eq_fillion_mean_theta}
\end{eqnarray}
Combining the results above with Eq.~\eqref{eq_mean_std_def} and using Table~\ref{table_residuals_Fillion} we obtain the corresponding two-sigma bounds shown in Fig.~\ref{fig_bounds_joint} in blue (radial) and yellow (polar) lines. Note that these results match our earlier estimate from Eq.~\eqref{eq_estimate}.

%%%%%%%%%%%%%%%%%%%%%%
\subsection{Satellite data} \label{sec_limit_satellite}
\indent

\begin{figure}[t!]
\centering
\begin{minipage}[b]{0.90\linewidth}
\includegraphics[width=\textwidth]{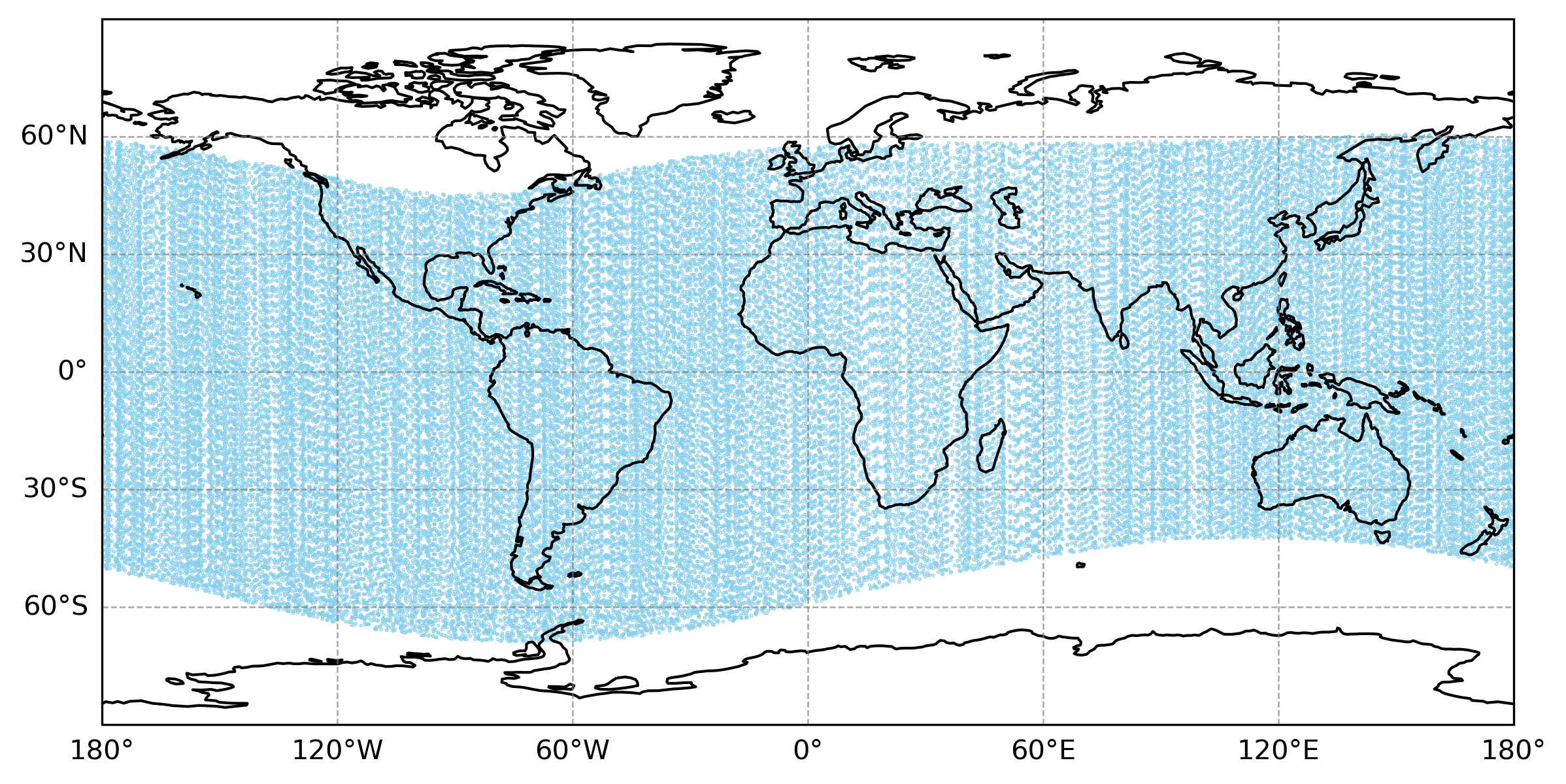}
\end{minipage} \hfill
\caption{Spatial coverage of Swarm~A during the years 2018-2020 limited to $\lambda_{\rm QD} < |55^\circ|$. }
\label{fig_map_sat}
\end{figure}

Geomagnetic data from satellite missions have provided high-precision vector measurements with unprecedented spatial coverage, cf. Fig.~\ref{fig_map_sat}. Since the introduction of satellites as platforms for measurements of the geomagnetic field -- the first high-resolution satellite was OGO-2, launched in 1965~\cite{Geo_sats_rev} -- much progress has been made: modern satellites collect vector magnetic data with sub-nT precision, contributing to significant refinement in global models~\cite{WMM2025, IGRF}. Furthermore, in the specific context of LSV, the tilted orbit relative to the equatorial plane means that instruments onboard of a satellite will in general be sensitive to combinations of the CFJ background that are otherwise inaccessible to ground observatories~\cite{Bluhm_2003}.

Finlay {\it et al.}~\cite{Finlay_2020} and Kloos {\it et al.}~\cite{Kloos} computed residuals between the CHAOS model and measurements from several satellites. The best agreement was achieved by the three identical Swarm satellites (A, B and C), launched in November 2013 into low-Earth orbits and providing the most precise measurements of the near-Earth geomagnetic field so far. Swarm~A and C fly side-by-side at an initial altitude $h = 450$~km and inclination $\zeta = 87.3^\circ$, whereas Swarm~B was initially placed at $h = 530$~km with $\zeta = 88^\circ$. The three spacecraft have prograde orbits with period $\tau_s = 94$~min; the associated orbital angular frequency is $\omega_s = 1.1 \times 10^{-3}$~rad/s~\cite{ref_litho, swarm, Visser}. Henceforth we will focus on Swarm~A; we expect similar results would be obtained with Swarm B or C.

Another important orbital element is the right ascension of the ascending node, the azimuthal angle $\alpha$ marking the intersection of the orbit in its ascending path with the equatorial plane. It is measured from the positive $X$-axis of the SCF, cf. App.~\ref{app_sat} and Fig.~2 of Ref.~\cite{Bluhm_2003}, and is not fixed since Earth is not perfectly spherical. In fact, it precesses at a rate~\cite{Brown_orbit} 
\begin{equation} \label{eq_nodal_drift_rate}
\omega_{\alpha} \approx -\frac{3}{2}  \frac{J_2 \, \omega_s \cos\zeta}{ \left[ 1 + (h/R_\oplus) \right]^2 } \, ,
\end{equation}
where $J_2 \approx 1.08 \times 10^{-3}$ is Earth's second dynamic form factor that characterizes the deviation of its shape from a perfect sphere. Plugging in the parameters for the Swarm-A satellite we find $\omega_{\alpha} \approx -7.4 \times 10^{-8}$~rad/s~\cite{Visser} corresponding to a period $\tau_\alpha = 2\pi/|\omega_{\alpha}| \approx 2.7$~yr. Therefore, the spatial components of the CFJ background\footnote{See Sec.~\ref{sec_limit_phi_k0_sat} for a discussion on the time component.} are modulated by two main frequencies, $\omega_s$ and $\omega_\alpha \ll \omega_s$, cf. App.~\ref{app_sat}.

\begin{table}[t!]
\centering
\begin{tabular}{|c|c|c|c|c|c|c|}
\hline
$\langle  \delta B_{r} \rangle$ & $\sigma_{\delta B_r}$ & $\langle  \delta B_{\theta} \rangle$ & $\sigma_{\delta B_\theta}$ & $\langle  \delta B_{\varphi} \rangle$ & $\sigma_{\delta B_\varphi}$  \\ \hline\hline
0.3 & 1.1 & 0.0 & 2.0 & -0.5 & 2.8 \\ \hline
\end{tabular}
\caption{Mean ($\langle  \delta B_{u} \rangle$) and standard deviation ($\sigma_{\delta B_u}$), with $u = \{ r, \theta, \varphi \}$, of the residuals between Swarm-A data and predictions of the CHAOS-8 and DIFI models. The initial $72257$ data points were filtered down by $\lesssim 0.2\%$ with the IQR method and weighted by the normalized area of the respective angular patches, cf. Eqs.~\eqref{eq_def_mean_residuals_obs} and~\eqref{eq_def_std_residuals_obs}. All values are given in nT.}\label{table_residuals_sat}
\end{table}

\begin{figure}[t!]
\centering
\begin{minipage}[b]{1.\linewidth}
\includegraphics[width=\textwidth]{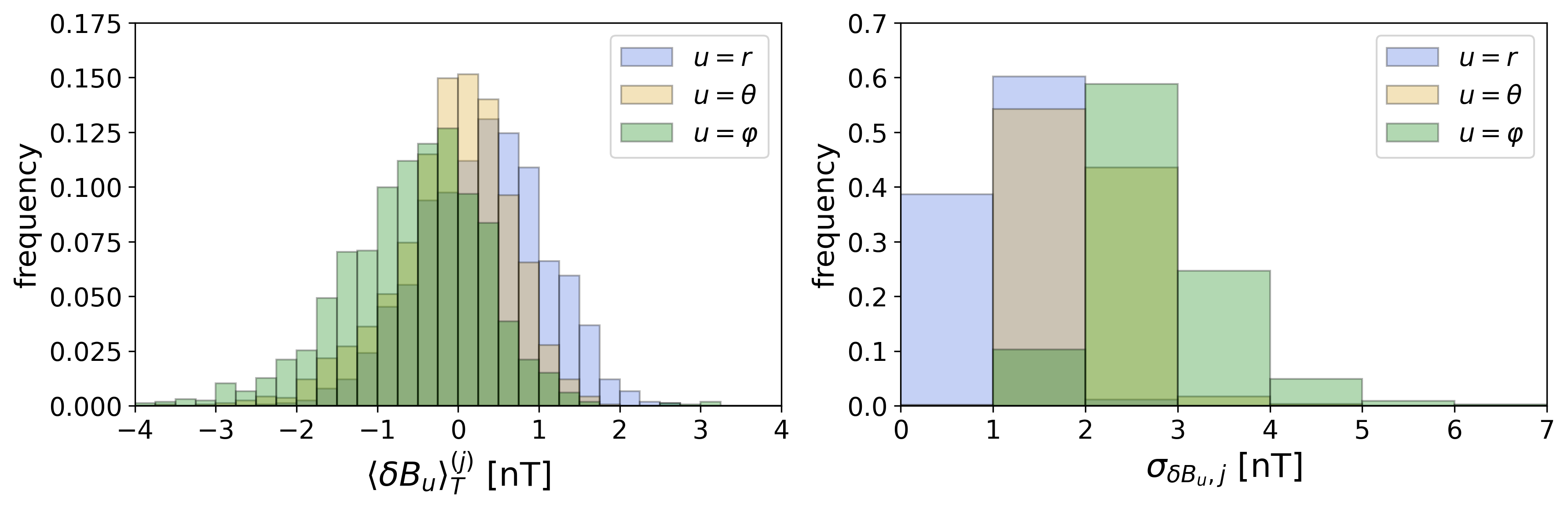}
\end{minipage} \hfill
\caption{Distributions of the filtered residuals in the non-empty angular patches of size $\Delta = 5^\circ$ following Eqs.~\eqref{eq_def_mean_residuals_T_obs} and~\eqref{eq_def_std_residuals_j_obs}. The height of each bin is the number of data in that bin divided by the total number; the sum of all heights is equal to one. }
\label{fig_hist_sat}
\end{figure}

Finlay {\it et al.} tested the CHAOS-7 model against satellite data in the period 1999-2020~\cite{Finlay_2020}. Particularly relevant are the results from the Swarm satellites, whose data are well matched by the model: the residuals for vector data have means $\lesssim 0.1$~nT with spreads $\lesssim 2.5$~nT. Besides the possibility of constraining different combinations of the CFJ 4-vector, if we compare these values with those for ground observatories in Table~\ref{table_residuals_Fillion}, we can naively expect an improvement in the bounds by a factor of $\lesssim 10$. Once again, we employ the version 8.3 of the CHAOS-8 model~\cite{Kloos, Kloss_Finlay_Olsen_Toffner-Clausen_Gillet_Grayver_2026} and the DIFI model~\cite{DIFI} and compare them with Swarm-A data collected every minute during nighttime between 1 and 5 hours local time between 18 March 2018, at 06:40 UTC, and 21 November 2020, at 10:23 UTC. Also, vector data are limited to $\lambda_{\rm QD} < |55^\circ|$, cf. Fig.~\ref{fig_map_sat}, and must satisfy ${\rm Hp30} < 1$~\cite{Hp30}. As done in Sec.~\ref{sec_limit_ground}, we remove $\approx 150$ large outliers from the initial $72257$ data points with the IQR method. Next, we follow the averaging procedure outlined in Sec.~\ref{sec_limit_ground}, where we split the globe into angular patches with sides of size $\Delta = 5^\circ$. Equations~\eqref{eq_def_mean_residuals_obs}-\eqref{eq_def_std_residuals_obs} remain valid, but Eq.~\eqref{eq_def_mean_cfj_obs} must be slightly modified -- this discussion will be postponed to Sec.~\ref{sec_limit_phi_k0_sat}, cf. Eq.~\eqref{eq_def_mean_cfj_sat}. Our results are summarized in Table~\ref{table_residuals_sat} and in Fig.~\ref{fig_hist_sat}. %Despite the smaller sample, we reach the same ballpark of Ref.~\cite{Finlay_2020}.

Assuming, as we did for ground observatories, that the frame in which satellite measurements are provided is aligned with the standard satellite reference frame as defined in App.~\ref{app_sat}, the spherical components of the CFJ magnetic field take the form of
Eqs.~\eqref{eq_B_radial_lab}-\eqref{eq_B_phi_lab}, but with $r \rightarrow r_s = R_\oplus + h$, $k_x \rightarrow k_\theta$, $k_y \rightarrow k_\varphi$ and $k_z \rightarrow k_r$ as given by Eqs.~\eqref{eq_app_kr_sat}-\eqref{eq_app_kphi_sat}. These expressions are given in terms of sines and cosines of the polar and azimuthal angles $\theta$ and $\varphi$ which describe the satellite's angular position relative to a non-rotating frame fixed to Earth's center with axes aligned to those of the SCF. Note that the azimuthal angle cannot be directly translated into geographic longitude. Apart from the special case of an equatorial orbit ($\zeta = 0$ or $\zeta = \pi$), both $\theta$ and $\varphi$ are explicitly time dependent via $\psi_s = \omega_s T_s$, the angle describing how far the satellite is along its orbit, and the slowly drifting position of the ascending node $\alpha = \alpha_0 + \omega_{\alpha}T_s$, cf. Eqs.~\eqref{eq_app_cos_theta}-\eqref{eq_app_cos_phi}. The relevant time coordinate is $T_s = T - T_{s,0}$, where $T_{s,0}$ may be chosen as any moment when the satellite crosses the equatorial plane in its ascending path, cf. App.~\ref{app_sat}, This complex time dependence of the CFJ background will allow other combinations of $\{ k_X, k_Y, k_Z \}$ to be accessed when compared to ground observatories.

\begin{figure}[t!]
\centering
\begin{minipage}[b]{0.8\linewidth}
\includegraphics[width=\textwidth]{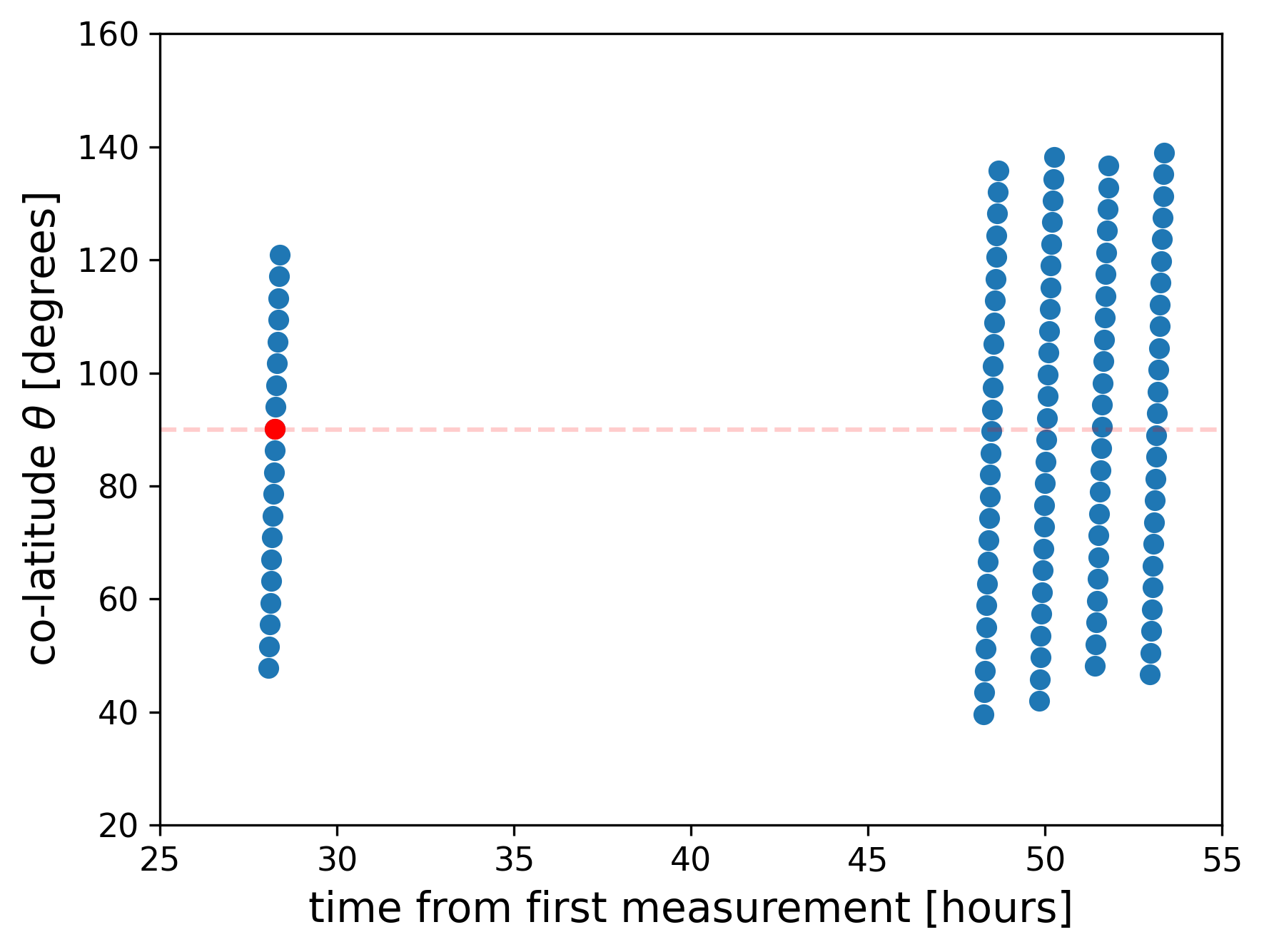}
\end{minipage} \hfill
\caption{Co-latitude of the Swarm-A satellite during the first few days of our dataset. The first data point was recorded on 18 March 2018, at 06:40 UTC. The first data point recorded at a time when the satellite crosses the equator ($\theta = \pi/2$), marked by the dashed red line, in its descending path towards the South Pole, is at $T_D =$~19 March 2018, at 10:55 UTC. This point is marked in red and its (geographic) longitude is $-90.5^\circ$. As discussed in the main text, $T_{s,0}$ can be placed at 19 March 2018, at 10:08 UTC -- not shown in the plot above -- at a (geographic) longitude $+89.5^\circ$. }
\label{fig_orbit}
\end{figure}

We need to determine a suitable $T_{s,0}$ from our data. Onboard of a satellite, the CFJ field depends on time through $\Psi = \Omega_\oplus T$, $\psi_s = \omega_s T_s$ and $\alpha = \alpha_0 + \omega_{\alpha}T_s$ with $T_s = T - T_{s,0}$. As discussed in App.~\ref{app_sat}, the time variable $T$ starts on the 2000 vernal equinox (20 March, at 07:35 UTC), but $T_{s,0}$ may be chosen as any moment where the satellite crosses the equatorial plane ($\theta = \pi/2$) in its ascending path. In Fig.~\ref{fig_orbit} we show the first four hours of data from the Swarm-A satellite starting on 18 March 2018, at 06:40 UTC. For this initial time period the only valid measurements were made when the satellite was in its descending path towards the South Pole. The first time it crosses the descending node is at $T_D =$~19 March 2018, at 10:55 UTC (at a geographic longitude of $-90.5^\circ$), but the orbit is symmetric around the line of nodes\footnote{At least to a very good approximation in the short time frame between two consecutive passes.} and we can calculate $T_{s,0}$ via $T_{s,0} = T_D \pm \tau_s/2$. We choose the minus sign, placing $T_{s,0}$ at 19 March 2018, at 10:08 UTC; its (geographic) longitude is $+89.5^\circ$. Finally, we note that it is impossible to determine the angle $\alpha_0$ associated with our choice for $T_{s,0}$ without a detailed reconstruction of the orbit -- more on this in Sec.~\ref{sec_limit_radial_polar_sat}.

%%%%%%%%%%%%%%%%%%%%%%
\subsubsection{First-order term} \label{sec_limit_phi_k0_sat}
\indent

\begin{figure}[t!]
\centering
\begin{minipage}[b]{0.80\linewidth}
\includegraphics[width=\textwidth]{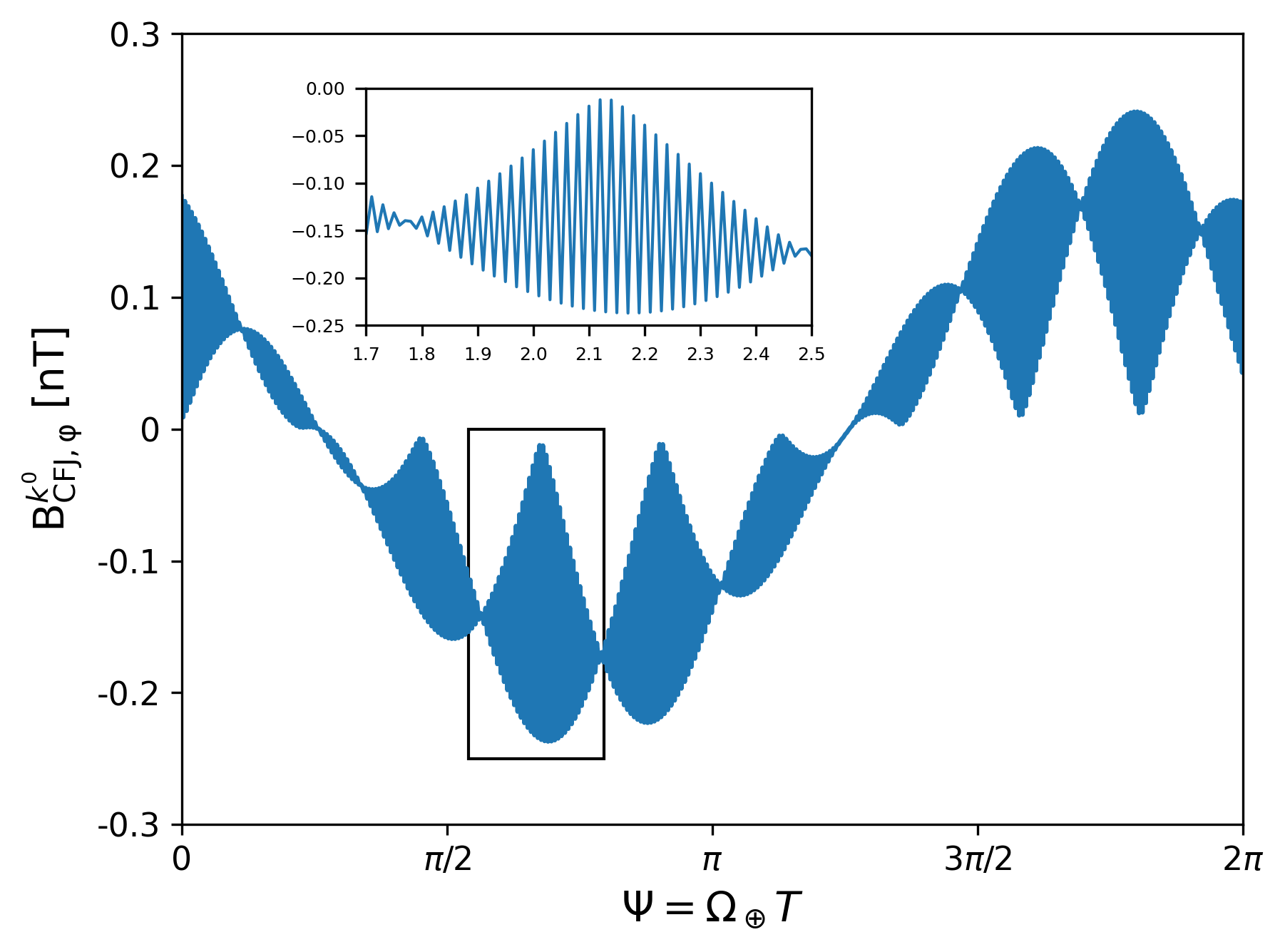}
\end{minipage} \hfill
\caption{Time variation of the first-order term of the CFJ azimuthal field, cf. Eq.~\eqref{eq_B_phi_sat_1}, during one year. Here $\zeta = 87.3^\circ$, $h = 450$~km and $k_X = 1$, $k_Y = 0.5$ and $k_Z = 1.5$ in units of $10^{-24} \, {\rm GeV}$. We also set $T_{s,0} = 0$ for simplicity. }
\label{fig_B_k0_sat}
\end{figure}

Let us start with the term linear in $k^0$. Setting $k \equiv \tilde{k} \times 10^{-24} \, {\rm GeV}$, we have
\begin{eqnarray} \label{eq_B_phi_sat_1}
{\rm B}_{\rm CFJ, \varphi}^{k^0} & = & -\frac{\mu \beta_\oplus \sqrt{1 - s_\zeta^2 s_{\psi_s}^2}}{2\pi (R_\oplus + h)^2} \left[ k_X s_\Psi - \left( c_\eta k_Y + s_\eta k_Z \right) c_\Psi \right]  \nonumber \\
& = & -\frac{\mu \beta_\oplus}{2\pi (R_\oplus + h)^2}
\displaystyle\sum_{l}^{X,Y,Z} c_{l} \, \tilde{k}_l \, ,
\end{eqnarray}
which is shown in Fig.~\ref{fig_B_k0_sat} for one sidereal day. The coefficients $c_{l}$ carrying the time dependence via $\psi_s = \omega_s T_s$ and $\Psi = \Omega_\oplus T$ are
\begin{equation} \label{eq_c_l}
c_X = s_\Psi \sqrt{1 - s_\zeta^2 s_{\psi_s}^2} \, , \quad c_Y = -c_\eta c_\Psi \sqrt{1 - s_\zeta^2 s_{\psi_s}^2}  \quad {\rm and} \quad c_Z = -s_\eta c_\Psi \sqrt{1 - s_\zeta^2 s_{\psi_s}^2} \, .
\end{equation}

\newpage

The averages are performed as outlined in Sec.~\ref{sec_limit_ground} by dividing the globe in angular patches of size $\Delta$ and weighing each patch by its (normalized) area. In particular, Eq.~\eqref{eq_def_mean_cfj_obs} must be modified to
\begin{equation} \label{eq_def_mean_cfj_sat}
\langle {\rm B}_{\rm CFJ, u} \rangle = \displaystyle\sum_{j}  w_j \, N_{j}^{-1} \displaystyle\sum_{{\rm i} \, {\rm in \, patch \, j}}  {\rm B}_{\rm CFJ, u}(R_\oplus + h_i; T_{s,i})  \, 
\end{equation}
with $T_{s,i}$ being related to the time stamps of the data in patch $j$ at altitude $h_i$. There are a few reasons for this change. First, for ground-based observers, the CFJ fields depend on a single frequency, $\Omega_\oplus$ for first-order terms and $\omega_\oplus$ for the second-order fields. Connected to this is the very long (1997-2024) span of hourly ground data, with which a numerical time averaging ends up being essentially equivalent to our analytical results, Eqs.~\eqref{eq_Br_avg_T} and~\eqref{eq_Btheta_avg_T}. In our present context of satellite-based measurements, though, the presence of two frequencies and the shorter span of Swarm-A data do not allow for the calculation of a simple analytical expression for the time averages.

We start by extracting the time stamp and the altitude from the data points in patch $j$: knowing $T_{s,0}$ we may use the time stamps to find $T_{s,i}$ for each measurement at $h_i$ in that patch. In this way we are able to calculate the mean CFJ field in each patch, which are then averaged over all non-empty patches according to Eq.~\eqref{eq_def_mean_cfj_sat}. Doing this, we find
\begin{equation} \label{eq_phi_k0_2} 
\langle {\rm B}_{\rm CFJ, \varphi}^{k^0} \rangle = \displaystyle\sum_{l}^{X,Y,Z} \langle b_{l} \rangle \, \tilde{k}_a \tilde{k}_b = -\left(4.3 \times 10^{-3} \, {\rm nT} \right) \left[ \tilde{k}_X + 0.02 \, \tilde{k}_Y + 0.01 \, \tilde{k}_Z \right]  \, .
\end{equation}
Together with the statistics on the residuals displayed in Table~\ref{table_residuals_sat}, this turns into the constraint
\begin{equation} \label{eq_k0_sat_data}
k_X + 0.02 \, k_Y + 0.01 \, k_Z = \left( 1.2 \pm 6.6 \right)  \times  10^{-22} \, {\rm GeV} \, ,
\end{equation}
which is better than the estimate~\eqref{eq_limit_k0_ground_CFJ}. The cause for the relative weakness of this bound will be discussed in Sec.~\ref{sec_sensis}, but two factors play a role. The first is the suppression due to $\beta_\oplus \approx 10^{-4}$. More importantly, though, in Fig.~\ref{fig_B_k0_sat} we see how the fast oscillations with period $\tau_s = 94$~min distort the overall yearly variation of $k^0$ with $\Omega_\oplus \ll \omega_s$, which itself averages to zero. We thus expect that $\langle {\rm B}_{\rm CFJ, \varphi}^{k^0} \rangle \rightarrow 0$ if the data covered a longer time period (see Sec.~\ref{sec_sensi_first_order_sat}).

%%%%%%%%%%%%%%%%%%%%%%
\subsubsection{Second-order fields} \label{sec_limit_radial_polar_sat}
\indent

\begin{figure}[t!]
\centering
\begin{minipage}[b]{0.80\linewidth}
\includegraphics[width=\textwidth]{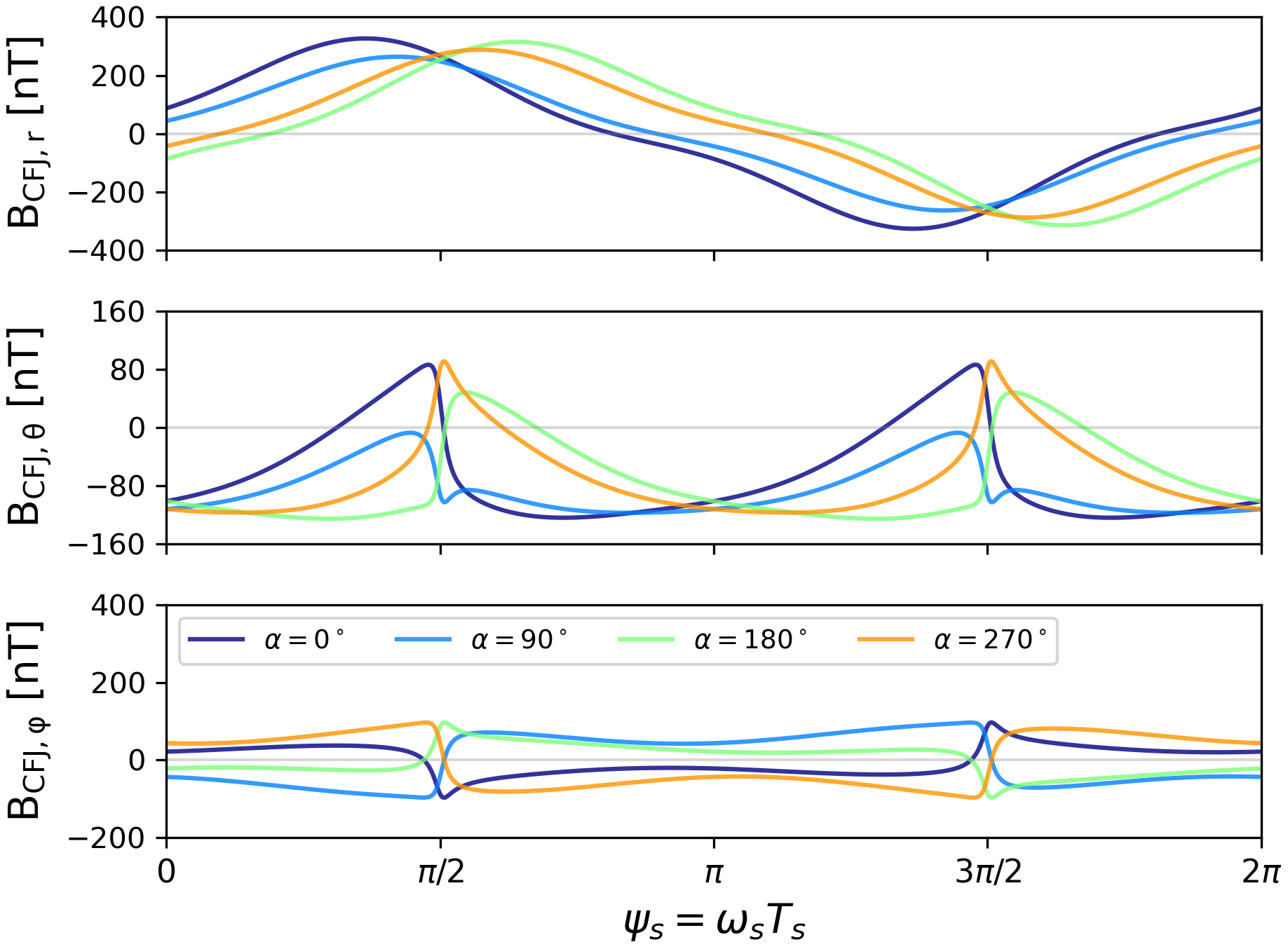}
\end{minipage} \hfill
\caption{Time variation of the radial, polar and azimuthal (only its quadratic part) components of the CFJ field, cf. Eqs.~\eqref{eq_B_rad_sat} and~\eqref{eq_B_theta_phi_sat}. We limit ourselves to one full revolution for different values of $\alpha$, which is essentially constant during this time frame. Here we set $\zeta = 87.3^\circ$, $h = 450$~km and $k_X = 1$, $k_Y = 0.5$ and $k_Z = 1.5$ in units of $10^{-24} \, {\rm GeV}$.}
\label{fig_B_time_sat}
\end{figure}

Let us move on to the second-order fields. The spherical components of the CFJ field take the form of
Eqs.~\eqref{eq_B_radial_lab}-\eqref{eq_B_phi_lab} with the substitutions $k_x \rightarrow k_\theta$, $k_y \rightarrow k_\varphi$ and $k_z \rightarrow k_r$, cf. Eqs.~\eqref{eq_app_kr_sat}-\eqref{eq_app_kphi_sat}; the resulting field components are illustrated in Fig.~\ref{fig_B_time_sat} for the parameters of the Swarm-A satellite during the period of one revolution. Due to the more involved structure of the $\{ k_r, k_\theta, k_\varphi \}$, the expressions for the field components become lengthier, so here we adopt a more compact notation. For the radial component we have
\begin{equation} \label{eq_B_rad_sat}
{\rm B}_{\rm CFJ, r} =  \frac{\mu}{2\pi (R_\oplus + h)} \displaystyle\sum_{a,b}^{X,Y,Z} c_{ab}^{(r)} \, k_a k_b \, ,
\end{equation}
which is shown in the upper panel of Fig.~\ref{fig_B_time_sat} for different choices of $\alpha$. The coefficients $c_{ab}^{(r)}$, functions of $\zeta$, $\alpha$ and $\psi_s$, are conveniently listed in Eqs.~\eqref{eq_app_c_XX_rad}-\eqref{eq_app_c_ZZ_rad} and shown in the bottom panel of  Fig.~\ref{fig_c_ij_all}.

\begin{figure}[t!]
\centering
\begin{minipage}[b]{0.80\linewidth}
\includegraphics[width=\textwidth]{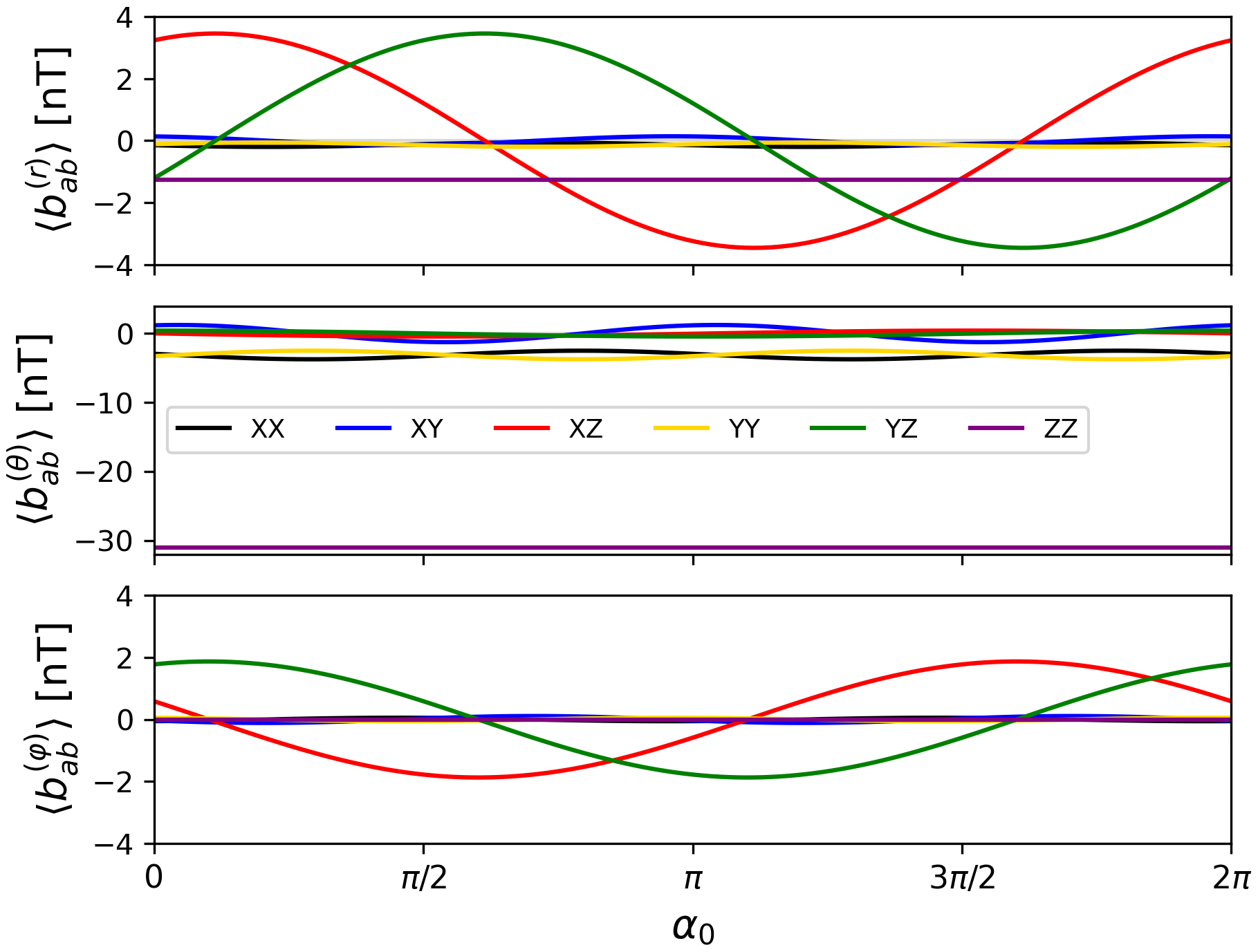}
\end{minipage} \hfill
\caption{Averaged coefficients of the components of the CFJ as functions of $\alpha_0$, cf. Eqs.~\eqref{eq_b_ij_phi} and~\eqref{eq_B_theta_phi_sat}. Note that $\langle b_{ZZ}^{(r)} \rangle$, $\langle b_{ZZ}^{(\theta)} \rangle$ and $\langle b_{ZZ}^{(\varphi)} \rangle$ are independent of $\alpha_0$.}
\label{fig_b_alpha_0}
\end{figure}

We must keep in mind that the initial angle $\alpha_0$ of the ascending node $\alpha = \alpha_0 + \omega_{\alpha}T_s$ is in principle unknown. Performing the average using the time-stamped data, we get
\begin{equation} \label{eq_b_ij_phi} 
\langle {\rm B}_{\rm CFJ,r} \rangle = \displaystyle\sum_{a,b}^{X,Y,Z} \langle b_{ab}^{(r)} \rangle \, \tilde{k}_a \tilde{k}_b   \, , 
\end{equation}
where the $\langle b_{ab}^{(r)} \rangle$ are effectively functions of $\alpha_0$. The polar and azimuthal components are:
\begin{equation} \label{eq_B_theta_phi_sat}
  \begin{split}
    {\rm B}_{\rm CFJ, \theta} & =  \frac{\mu}{8\pi (R_\oplus + h)}  \displaystyle\sum_{a,b}^{X,Y,Z} c_{ab}^{(\theta)} \, k_a k_b   \\
{\rm B}_{\rm CFJ, \varphi}^{k^2} &= -\frac{\mu}{4\pi (R_\oplus + h)} \displaystyle\sum_{a,b}^{X,Y,Z} c_{ab}^{(\varphi)} \, k_a k_b
  \end{split}
\quad\rightarrow\quad
  \begin{split}
    \langle {\rm B}_{\rm CFJ, \theta} \rangle &= \displaystyle\sum_{a,b}^{X,Y,Z} \langle b_{ab}^{(\theta)} \rangle \, \tilde{k}_a \tilde{k}_b \, ,  \\
    \langle {\rm B}_{\rm CFJ, \varphi}^{k^2} \rangle &= \displaystyle\sum_{a,b}^{X,Y,Z} \langle b_{ab}^{(\varphi)} \rangle \, \tilde{k}_a \tilde{k}_b \, .
  \end{split}
\end{equation}
The fields are shown in Fig.~\ref{fig_B_time_sat} for different choices of $\alpha$, $c_{ab}^{(\theta)}$ and $c_{ab}^{(\varphi)}$ are listed in Eqs.~\eqref{eq_app_c_XX_theta}-\eqref{eq_app_c_ZZ_theta} and in Eqs.~\eqref{eq_app_c_XX_phi}-\eqref{eq_app_c_ZZ_phi}, respectively (see top and middle panels of Fig.~\ref{fig_c_ij_all}). The $\langle b_{ab}^{(u)} \rangle$ are shown in Fig.~\ref{fig_b_alpha_0}.

\begin{figure}[t!]
\centering
\begin{minipage}[b]{0.9\linewidth}
\includegraphics[width=\textwidth]{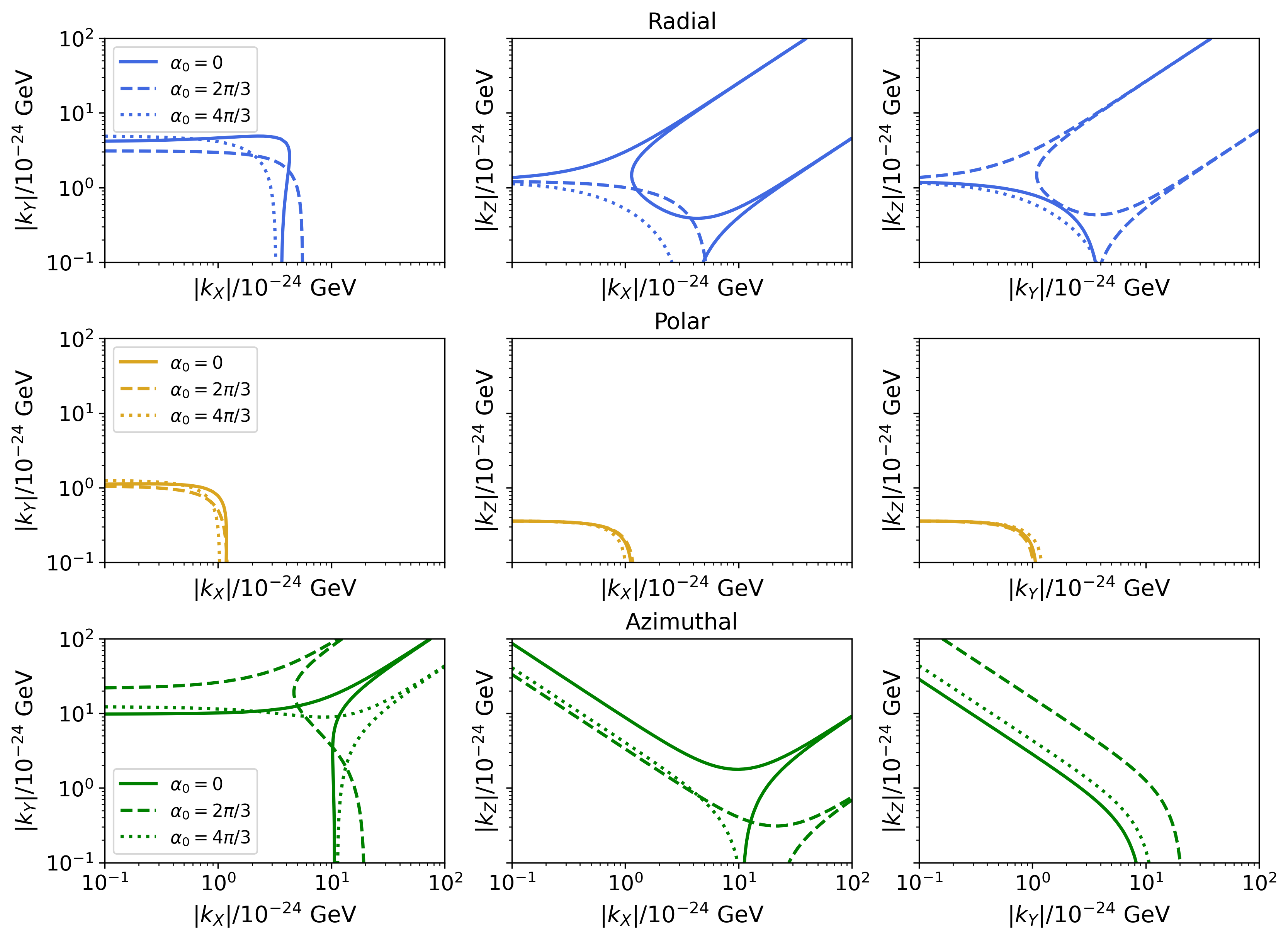}
\end{minipage} \hfill
\caption{Two-sigma bounds on combinations of the CFJ background 4-vector in the SCF derived from Swarm-A data (2018-2020). In blue, yellow and green are the bounds from the radial, polar and azimuthal components, respectively. In each plot we show the bounds resulting from Eqs.~\eqref{eq_b_ij_phi} and~\eqref{eq_B_theta_phi_sat} for three different (arbitrary) choices of the initial angle $\alpha_0$, cf. Fig.~\ref{fig_b_alpha_0}, while setting the remaining $k_i$ to zero (e.g., $k_Z = 0$ in the left column). The regions of parameter space within the curves are allowed.}
\label{fig_bounds_alpha0}
\end{figure}

As illustrated in Fig.~\ref{fig_b_alpha_0}, the fixed, but unknown initial angle $\alpha_0$ has a potentially strong impact on the averaged components of the second-order CFJ fields. This is only so because, neglecting $\tau_s \ll \tau_\alpha$, the time span of the available data, $\Delta t \approx 2.7$~yr, is not exactly an integer multiple of $\tau_\alpha$, thus causing the (time) average to depend on the constant phase $\alpha_0$. This dependence is enhanced by the fact that $\Delta t$ and $\tau_\alpha$ are of similar sizes, that is, we do not have $\Delta t \gg \tau_\alpha$. These issues will be discussed in further detail in Sec.~\ref{sec_sensis}, but for now we keep $\alpha_0$ as a free parameter. The residuals have two-sigma spreads $\lesssim 6$~nT, whereas most of the coefficients $\langle b_{ab} \rangle$ displayed in Fig.~\ref{fig_b_alpha_0} have relatively small amplitudes, at the range of $\lesssim 0.1$~nT. These numbers suggest that the bounds will be mostly at the level of $\approx 10^{-24}-10^{-23} \, {\rm GeV}$ for the individual $\{ k_X, k_Y, k_Z \}$. In Fig.~\ref{fig_bounds_alpha0} we show the actual two-sigma bounds obtained by combining the residuals between Swarm-A data and the predictions from the CHAOS-8 and DIFI models as collected in Table~\ref{table_residuals_sat} with the averaged fields from Eqs.~\eqref{eq_b_ij_phi} and~\eqref{eq_B_theta_phi_sat} for different choices of $\alpha_0$. The impact of the value of $\alpha_0$ is large, in particular for the bounds from the radial and azimuthal components.

One exception is worth mentioning: the polar component of the CFJ field. For it we find that $\langle b_{ZZ}^{(\theta)} \rangle = -31.1$~nT, a constant, while all other averaged coefficients lie much closer to zero oscillating with small amplitudes, cf. Fig.~\ref{fig_b_alpha_0}. This explains why $|k_Z| \approx 4 \times 10^{-25} \, {\rm GeV}$ for $|k_X|, |k_Y| \lesssim 10^{-24} \, {\rm GeV}$ is our best bound, as can be seen in the middle row of Fig.~\ref{fig_bounds_alpha0}. In contrast, for the azimuthal component all $\langle b_{ab}^{(\varphi)} \rangle$ vary around zero with generally small amplitudes, accounting for the overall lack of improvement of the bounds relative to those based on ground data with much larger spreads, cf. Fig.~\ref{fig_bounds_joint}, even though the averaged coefficients $\langle b_{XZ}^{(\varphi)} \rangle$ and $\langle b_{YZ}^{(\varphi)} \rangle$ may vary up to $\approx 1.9$~nT for specific choices of $\alpha_0$. Similar arguments hold for the bounds obtained from the radial component of the CFJ field.

%%%%%%%%%%%%%%%%%%%%%%%%%%%%
\section{Projected sensitivities} \label{sec_sensis}
\indent 

The bounds shown in Fig.~\ref{fig_bounds_alpha0} are our final results with the data available. However, as remarked in the previous section, the fact that the time span of the data, $\Delta t$, is similar, but not identical, to the largest period dictating the time variation of the CFJ field as measured by a satellite, $\tau_\alpha$, allows the time averages to depend on the angle $\alpha_0$ between the ascending node and the $X$-axis of the SCF at $T_s = T_{s,0}$ (see App.~\ref{app_sat}). This angle can only be determined by a detailed reconstruction of the orbit, which is beyond the scope of this work.

We have shown that, depending on the (arbitrary) choice of $\alpha_0$, individual components of the CFJ background may be constrained at the level of $ 10^{-25} \, {\rm GeV}$, a significant improvement over the bounds from ground data, cf. Fig.~\ref{fig_bounds_joint}. This motivates the question of whether similarly precise satellite measurements of the geomagnetic field could lead to even tighter constraints if conducted over time frames much longer than the roughly three years or if the satellite's orbit had a different inclination $\zeta$. The altitude $h$ could also be varied: from our initial estimate~\eqref{eq_estimate}, it is clear that lower orbits are advantageous. Very-low Earth orbits have been investigated due to a few promising aspects for Earth-observation missions, such as improved spatial resolution of imagery and accuracy in geospatial positioning. There are, however, several practical challenges that significantly limit the mission lifetime: the need for constant propulsion to counteract aerodynamic drag, spacecraft degradation due to the interaction with atmospheric gases, among others; for a detailed review, see Ref.~\cite{VLEO}. Hence, for the sake of definitiveness we henceforth fix the altitude at $h = 450$~km.

The main issue with the bounds shown in Fig.~\ref{fig_b_alpha_0} is that they depend on $\alpha_0$. The time evolution of the $c_{ab}^{(u)} = c_{ab}^{(u)}(T_s)$, with $u = \{ r, \theta, \varphi\}$, depends on two periods, $\tau_s$ and $\tau_\alpha \gg \tau_s$, which are rational numbers. The overall period $\mathcal{T}$, defined such that $c_{ab}^{(u)}(T_s + \mathcal{T}) = c_{ab}^{(u)}(T_s)$, is given by $\mathcal{T} = n_s \tau_s = n_\alpha \tau_\alpha$ with integer $n_s, n_\alpha$: in our case $\tau_\alpha/\tau_s = n_s/n_\alpha \approx 550000/37$ and $\mathcal{T} \approx 99.6$~yr. Our objective is to average $c_{ab}^{(u)}$ over a time period $\Delta t$; here we omit the angular weighting for brevity. Taking the continuous case for simplicity, the time average is
\begin{eqnarray} \label{eq_time_av_def}
\langle c_{ab}^{(u)} \rangle_{T} = \frac{1}{\Delta t} \displaystyle\int_0^{\Delta t} c_{ab}^{(u)}(T_s) \, dT_s \, ,
\end{eqnarray}
but we know that $\Delta t$ will never be exactly equal to an integer multiple of $\mathcal{T}$, so let us write it as $\Delta t = n \mathcal{T} + \delta t$, where $n$ is an integer and $|\delta t| < \mathcal{T}$ is the mismatch. Expanding the integration domain of Eq.~\eqref{eq_time_av_def} in $n$ $\mathcal{T}$-periods, using $c_{ab}^{(u)}(T_s + \mathcal{T}) = c_{ab}^{(u)}(T_s)$ and neglecting $\delta t \ll n \mathcal{T}$ in the denominators, we find
\begin{eqnarray} \label{eq_time_av_def_2}
\langle c_{ab}^{(u)} \rangle_{T} = \langle c_{ab}^{(u)} \rangle_{\mathcal{T}} + 
\frac{1}{n \mathcal{T}} \displaystyle\int_0^{\delta t} c_{ab}^{(u)}(T_s) \, dT_s \, .
\end{eqnarray}

The coefficients $c_{ab}^{(u)}(T_s)$ are continuous functions of $T_s$ and, most importantly, are bounded, meaning that there exists a finite real number $M$ such that $|c_{ab}^{(u)}(T_s)| \leq M$ for any $T_s$. The direct effect of the $c_{ab}^{(u)}(T_s)$ being bounded is that the there is an upper limit for the integral appearing in the second term of Eq.~\eqref{eq_time_av_def_2}, namely, $M \delta t$. Putting all together, we obtain 
\begin{eqnarray} \label{eq_time_av_def_3}
\langle c_{ab}^{(u)} \rangle_{T} - \langle c_{ab}^{(u)} \rangle_{\mathcal{T}} \leq 
\frac{M \delta t}{n \mathcal{T}}  \, .
\end{eqnarray}
Now, if $M \delta t \ll n \mathcal{T}$, we do not need to bother about $\delta t$ and $\langle c_{ab}^{(u)} \rangle_{T} \approx \langle c_{ab}^{(u)} \rangle_{\mathcal{T}}$, whereby $\langle c_{ab}^{(u)} \rangle_{\mathcal{T}}$ is by definition independent of any phase constant. Therefore, for long observation times both $\delta t$ and $\alpha_0$ (and the choice of $T_{s,0}$) become irrelevant.

From Fig.~\ref{fig_c_ij_all} we notice that most of the $c_{ab}^{(u)}$ oscillate around zero, therefore we expect that the averages of these coefficients will tend to zero over observation times longer than the period considered in Sec.~\ref{sec_limit_satellite}. We wish to determine which of the $c_{ab}^{(u)}$ will have non-zero averages for longer observation times $\Delta t$, thus providing projections of the limits that could be attained if data were collected during much longer time scales. In order to achieve this goal we refrain from simulating new data, as this would require detailed knowledge of the satellite's orbit. Instead, we take the first filtered satellite measurement at time $t_i$ and note in which angular patch (of size $\Delta$) it falls. Next, we identify the last data point\footnote{This measurement was made just $3.3$~days prior to the overall final data point of our dataset, so the reduced dataset has roughly the same size as the original one.} located within that initial angular patch with time stamp $t_c$ and we cut the original dataset at this point. This reduced dataset is copied and, after shifting all time stamps by $t_c - t_i + 1 \, {\rm min}$, it is concatenated to the first reduced dataset. The extra minute is added to avoid having two data points with the same time stamp.

Repeating this procedure allows us to extend our original dataset to a longer time span with the advantage of automatically fulfilling all constraints, such as the limitation to $\lambda_{\rm QD} < |55^\circ|$. The extended dataset is as continuous as the original one, since contiguous data are separated by one minute intervals and the gaps caused by the aforementioned cuts (see Fig.~\ref{fig_orbit}) are also reproduced. For concreteness, we perform $80$ concatenations of the reduced dataset, corresponding to $\Delta t \approx 210$~yr of synthetic data. The average is performed according to Eq.~\eqref{eq_def_mean_cfj_sat}.

%%%%%%%%%%%%%%%%%%%%%%
\subsection{First-order term} \label{sec_sensi_first_order_sat}
\indent

Similar to the second-order terms, we have two different frequencies: the orbital frequency $\omega_s$ encoded in $\psi_s = \omega_s T_s$ and Earth's translational frequency $\Omega_\oplus \ll \omega_s$ appearing in $\Psi = \Omega_\oplus T$. However, different from the second-order terms, there is no unknown phase analogous to $\alpha_0$, because the origin of the time variables $T$ and $T_s$ are known. Performing the average of Eq.~\eqref{eq_B_phi_sat_1} according to Eq.~\eqref{eq_def_mean_cfj_sat} for the extended dataset as discussed above (with $h = 450$~km) we find that the overall pre-factor of $\langle {\rm B}_{\rm CFJ, \varphi}^{k^0} \rangle$, cf. Eq.~\eqref{eq_phi_k0_2}, is  $\approx 4 \times 10^{-5}$~nT for $\Delta t \approx 210$~yr. Clearly, increasing the observation time by a factor of 80 has reduced the amplitude of $\langle {\rm B}_{\rm CFJ, \varphi}^{k^0} \rangle$ by roughly the same amount. We thus expect that $\langle {\rm B}_{\rm CFJ, \varphi}^{k^0} \rangle \rightarrow 0$ in the long run, so no useful projections may be derived from the first-order azimuthal field.

%%%%%%%%%%%%%%%%%%%%%%
\subsection{Second-order fields} \label{sec_sensi_second_order_sat}
\indent

The coefficients $c_{ab}^{(\varphi)}$ are listed in App.~\ref{app_sat} and their (indirect) time dependence is shown in Fig.~\ref{fig_c_ij_all}. For $\Delta t \approx 210$~yr we find $\langle b^{(\varphi)}_{ab} \rangle \lesssim 4 \times 10^{-3}$~nT, a couple orders of magnitude smaller than for $\Delta t \approx 2.7$~yr, cf. Fig.~\ref{fig_b_alpha_0}. Therefore, $\langle {\rm B}_{\rm CFJ, \varphi} \rangle \rightarrow 0$ for long observation times; a similar conclusion is reached for $\langle {\rm B}_{\rm CFJ, r} \rangle$. For the polar component we find $\langle b^{(\theta)}_{ZZ} \rangle \approx 31 \, {\rm nT} \approx 10\langle b^{(\theta)}_{XX,YY} \rangle$ with the other $\langle b^{(\theta)}_{ab} \rangle$ being at least ten times smaller than the $\langle b^{(\theta)}_{XX,YY} \rangle$. Therefore, we have
\begin{equation} \label{eq_mean_theta_sat}
\langle {\rm B}_{\rm CFJ, \theta} \rangle \approx -3.1 \, {\rm nT} \left( \tilde{k}_X^2 + \tilde{k}_Y^2 + 10.0 \, \tilde{k}_Z^2  \right) \, 
\end{equation}
with a much milder dependence on $\alpha_0$ when compared to Fig.~\ref{fig_b_alpha_0}.

\begin{figure}[t!]
\centering
\begin{minipage}[b]{0.80\linewidth}
\includegraphics[width=\textwidth]{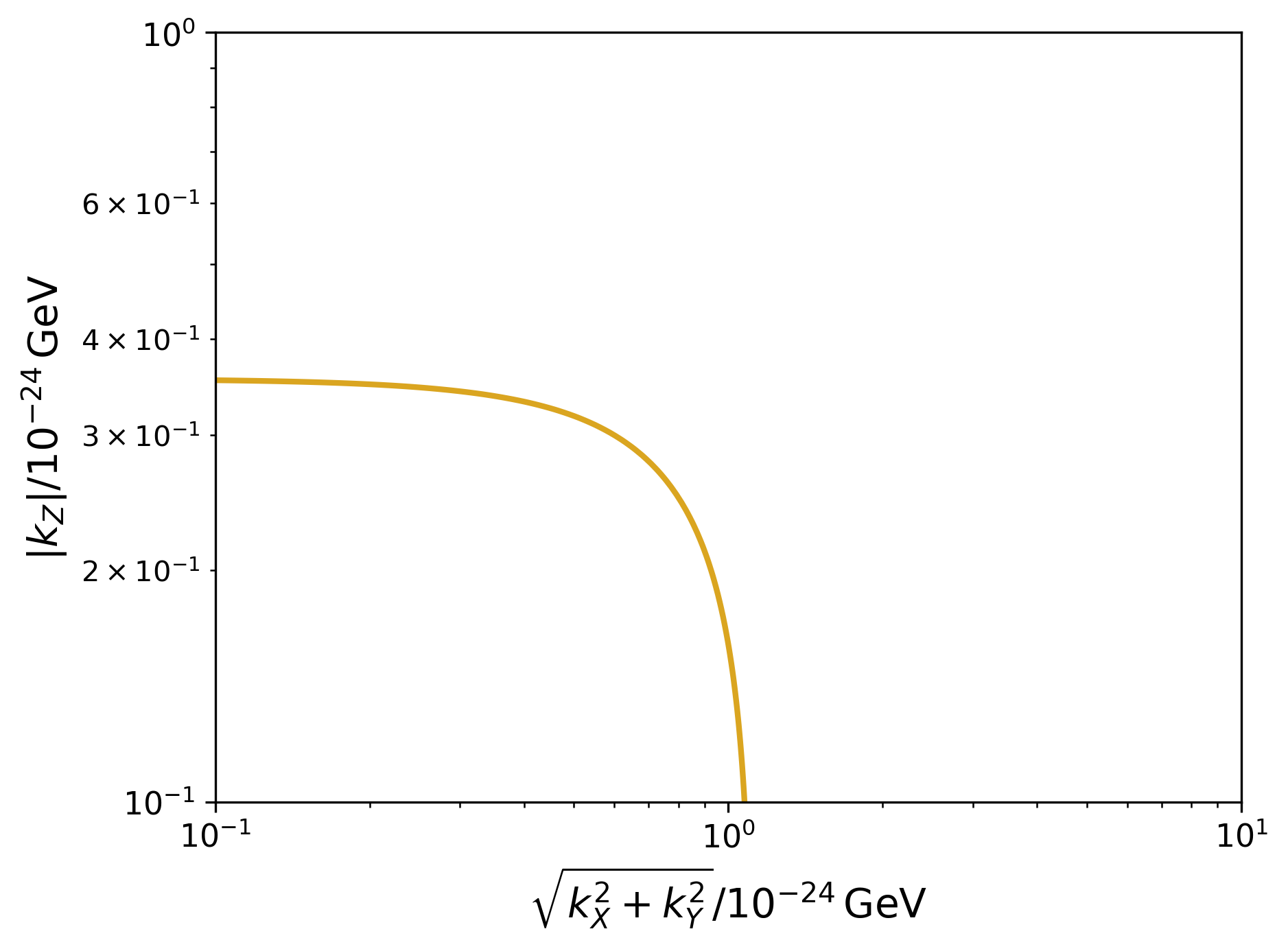}
\end{minipage} \hfill
\caption{Projected two-sigma bound on combinations of the CFJ background in the SCF for long-term satellite measurements of the polar component, cf. Eq.~\eqref{eq_mean_theta_sat}. The region of parameter space within the curve is allowed.}
\label{fig_sens_joint}
\end{figure}

The residuals of the polar component from roughly 20 years of Swarm-A data have means and standard deviations of $\approx 0.1$~nT and $\approx 2$~nT, respectively~\cite{Finlay_2020, Kloos}, in line with the shorter-term results from Sec.~\ref{sec_limit_satellite}, cf. Table~\ref{table_residuals_sat}. We take these values as benchmarks under the assumption that future improvements in modelling are unlikely to significantly reduce the spread of the residual differences between satellite data and geomagnetic models. Comparing this to Eq.~\eqref{eq_mean_theta_sat} we obtain the following one-sigma constraint:
\begin{equation} \label{eq_limit_theta_sat}
k_X^2 + k_Y^2 + 10.0 \, k_Z^2 \approx \left( 0.0 \pm 0.7 \right)  \times  10^{-48} \, {\rm GeV}^2 \, .
\end{equation}
The corresponding two-sigma projection is shown in Fig.~\ref{fig_sens_joint}. This result is very similar to those from actual satellite data shown in the middle row, central and right-hand plots, of Fig.~\eqref{fig_bounds_alpha0}. It is noteworthy that the long-term average of the polar component of the CFJ field displays the same combinations $k_X^2 + k_Y^2$ and $k_Z^2$ for an observer fixed to Earth's surface, cf. Eq.~\eqref{eq_Btheta_avg_T}, and onboard of a satellite.

%%%%%%%%%%%%%%%%%%%%%%
\subsection{Effect of the orbital inclination on the sensitivities} \label{sec_sensi_zeta}
\indent 

We have shown that the long-term averages of the CFJ field tend to zero, except for $\langle {\rm B}_{\rm CFJ, \theta} \rangle$, for which only the ``diagonal" combinations $k_X^2 + k_Y^2$ and $k_Z^2$ survive with meaningful amplitudes, cf. Eq.~\eqref{eq_mean_theta_sat}. To test if this is a consequence of the near-polar orbit ($\zeta = 87.3^\circ$) of Swarm~A, we evaluate the long-term time averages of the $b_{l}$ and $b_{ab}^{(u)}$, here respectively denoted by $\langle b_{l} \rangle_T$ and $\langle b_{ab}^{(u)} \rangle_T$, for different choices of $\zeta$. In order to keep the discussion simple and focus on the essential physics, we make a few assumptions. First, we use the extended dataset -- actually, only the time stamps. Second, we fix $h = 450$~km and $\alpha_0 = 0$. Finally, since changing $\zeta$ will impact the range of co-latitudes accessible to the satellite, the number of angular patches covered will be reduced: for $\zeta \lesssim 60^\circ$, equivalent to $30^\circ \lesssim \theta \lesssim 150^\circ$, the spatial coverage delimited by the requirement $\lambda_{\rm QD} < |55^\circ|$ will be diminished (see Fig.~\ref{fig_map_sat}). We then assume that enough angular patches are covered and that the time averages in Eq.~\eqref{eq_def_mean_cfj_sat} are independent of the angular patch $j$. Moreover, changing $\zeta$ affects $\omega_\alpha$, cf. Eq.~\eqref{eq_nodal_drift_rate}.

\begin{figure}[t!]
\centering
\begin{minipage}[b]{0.80\linewidth}
\includegraphics[width=\textwidth]{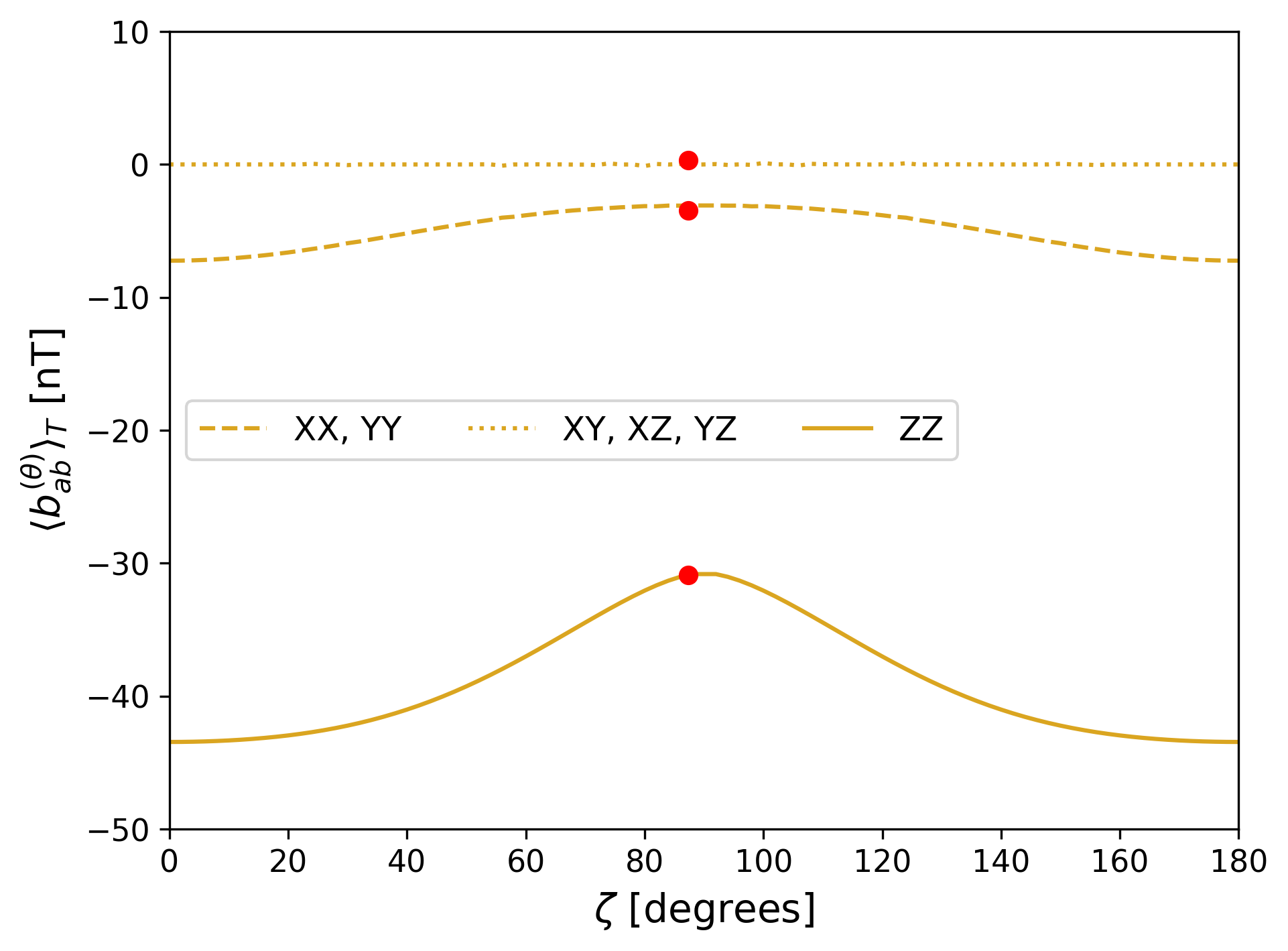}
\end{minipage} \hfill
\caption{ Long-term time-averaged coefficients $\langle b_{ab}^{(\theta)} \rangle_{T}$ of the polar component of the CFJ field as a function of the orbital inclination angle $\zeta$ relative to the equatorial plane. The red dots mark the values obtained for an orbit similar to that of the Swarm-A satellite with $h = 450$~km and $\zeta = 87.3^\circ$, namely, $10 \,\langle b_{XX}^{(\theta)} \rangle_T \approx 10 \,\langle b_{YY}^{(\theta)} \rangle_T \approx \langle b_{ZZ}^{(\theta)} \rangle_T \approx -31$~nT. See also the discussion in Sec.~\ref{sec_sensi_second_order_sat}.}
\label{fig_b_ij_zeta}
\end{figure}

Starting with the first-order terms from the azimuthal component of the CFJ field, we find $\langle b_l \rangle_T \approx 10^{-5} \, {\rm nT}$ for different values of $\zeta$, cf. Eq.~\eqref{eq_phi_k0_2}. This matches the value reported in Sec.~\ref{sec_sensi_first_order_sat}, indicating that changing $\zeta$ has little impact on the sensitivities of this term, which remain uninteresting as a source of competitive bounds on the CFJ background. Moving on to the second-order terms, we observe that $\langle b_{ab}^{(r)} \rangle_T$ and $\langle b_{ab}^{(\varphi)} \rangle_T$ are at the level of $\lesssim 10^{-2}$~nT\footnote{The fact that the $\langle b_{ab}^{(r)} \rangle_T$ and $\langle b_{ab}^{(\varphi)} \rangle_T$ are larger than the corresponding coefficients found in Sec.~\ref{sec_sensi_second_order_sat} is explained by the simplifications made in this section, in particular the angular averaging, which is ignored here. Nonetheless, the conclusions are maintained: the long-term averages of the radial and azimuthal components are much smaller than that of the polar component, independently of the orbital inclination $\zeta$.} with little absolute enhancement for different choices of $\zeta$. The polar component, on the other hand, is again more interesting. In Fig.~\ref{fig_b_ij_zeta} we show the variation of the $\langle b_{ab}^{(\theta)} \rangle$ with $\zeta$. The ``off-diagonal" terms are very close to zero, but the ``diagonal" ones have larger magnitudes varying by $\approx 30\%$ with $\zeta$. Overall, we conclude that $\zeta$, keeping $h$ and $\tau_s$ fixed, does not change which $\langle b_{ab}^{(u)} \rangle_T$ are finite or not in the long term.

Another interesting aspect of Fig.~\ref{fig_b_ij_zeta} is that nearly equatorial orbits with $\zeta \approx 0$ (prograde) or $\zeta \approx \pi$ (retrograde) would slightly enhance the sensitivity to time-averaged LSV effects. Such orbits are typically employed by geostationary satellites orbiting at $h = 35786$~km, but this is not strictly necessary and low-Earth orbits are possible\footnote{Satellites such as Lemur-2 Joel (USA) or Astrosat (India) operate in low-Earth, near-equatorial orbits with $h \sim 400-600$~km, though most satellites at low inclinations are geostationary. For a comprehensive database of currently active satellites, including their orbital parameters, see Ref.~\cite{list_sats}.}. For an equatorial orbit we have a partial equivalence with the case of a ground observatory on the equator ($\theta = \pi/2$) for which $\langle {\rm B}_{\rm CFJ, r} \rangle_{T} = 0$, cf. Eq.~\eqref{eq_Br_avg_T}, and $\langle {\rm B}_{\rm CFJ, \theta} \rangle_{T} \sim (k_X^2 + k_Y^2) + 6 \, k_Z^2$, cf. Eq.~\eqref{eq_Btheta_avg_T}. For a satellite with $\zeta \approx 0$ or $\zeta \approx \pi$, from Fig.~\ref{fig_b_ij_zeta} we have $\langle b_{XX}^{(\theta)} \rangle_{T} = \langle b_{YY}^{(\theta)} \rangle_{T} = -7.2$~nT and $\langle b_{ZZ}^{(\theta)} \rangle_{T} = -43.4$~nT, which are in the same proportion as in the case of a ground observatory. Unfortunately, near-equatorial orbits are impractical for studies of the geomagnetic field, since by definition they do not allow for a comprehensive enough spatial coverage.

Finally, let us address the long-term permanence of the ``diagonal" combinations $k_X^2 + k_Y^2$ and $k_Z^2$. Consider a point in the orbit at fixed polar angle $\theta$. During short observation times $\Delta t \lesssim \tau_\alpha$ the ascending node -- and the orbital plane -- does not have enough time to significantly precess. Consequently, for this point, its azimuthal angle $\varphi$ relative to the $X$-axis of the SCF will remain practically fixed. The short-term average singles out the only special direction available in this time frame, namely, the direction along the rotation axis of the satellite, so it is not surprising that the averages are in general non-zero.

Now, for $\Delta t \gg \tau_\alpha$ the orbital plane is able to precess and complete several full revolutions. Since the orbital plane slowly precesses, the azimuthal coordinate $\varphi$ of this point, which is measured by its projection onto the equatorial plane and relative to the $X$-axis of the SCF, will also change in time and end up describing a trajectory parallel to the equatorial plane in the SCF. Therefore, the long-term drift of the nodal line singles out the ``diagonal" combinations of the $\{ k_X, k_Y, k_Z \}$ that are characteristic of the equatorial plane, parallel to the $X-Y$-plane of the SCF, and the $Z$-axis perpendicular to it. This situation is analogous to that of a ground observatory at a fixed co-latitude whose azimuthal angle is given by $\psi_\oplus = \omega_\oplus T_\oplus$ -- over long times the averages also select the combinations $k_X^2 + k_Y^2$ and $k_Z^2$, cf. Eqs.~\eqref{eq_Br_avg_T} and~\eqref{eq_Btheta_avg_T}.

%%%%%%%%%%%%%%%%%%%%%%%%%%%%
\section{Concluding remarks} \label{sec_conc}
\indent

In this paper we calculated the magnetic field produced by a point-like magnetic dipole in the context of the CFJ electrodynamics. The magnetic field was obtained via the Green's method for a static source and displays, besides the standard Maxwellian dipolar field, contributions of first and second order in the spatial components of the CFJ background; higher orders are expected to be negligible and were not included.

The CFJ magnetic field alters the simple features of the standard Maxwellian dipolar field by adding novel azimuthal components, as well as extra terms acting to modify the radial and polar components. Modern global magnetic models indicate that the geomagnetic field is dominated by a dipolar field originating in Earth's liquid-iron outer core. Thus, if LSV is realized through the CFJ term~\eqref{eq_CFJ}, we can expect small, but potentially measurable, distortions to the geomagnetic field as modelled with standard Maxwellian electrodynamics. The CFJ vector is fixed in the approximately inertial frame centered at the Sun, but it will appear to periodically vary with characteristic frequencies (and harmonics thereof), sidereal and/or annual, if the observer is fixed to the terrestrial surface, or much shorter for instruments mounted on a satellite.

We focused on finite offsets that would worsen the agreement between the predictions of models and geomagnetic data. Our best bounds were obtained by analyzing the time- and angle-averaged differences between model and measurement (residuals) from roughly three years of Swarm-A data. In particular, the polar component of the field yields $|k_Z| \lesssim 4 \times 10^{-25} \, {\rm GeV}$ for $|k_X|, |k_Y| \lesssim 10^{-24} \, {\rm GeV}$ at the two-sigma level, cf. Fig.~\ref{fig_bounds_alpha0}. This represents an improvement of $\approx 10^4$ over previous bounds based on terrestrial phenomena~\cite{Mewes_2008, Kosteleck_2011}. Changing the inclination of the orbit has limited impact on the bounds, which depend on $k_X^2 + k_Y^2$ and $k_Z^2$ for long observation times. Improvements may be potentially attained if we pursue a time-resolved analysis such as done in Refs.~\cite{Arza_2022, Fedderke_2021, Sulai_2023} in the context of dark matter. The objective would be to filter out the LSV signal with its typical frequencies from geomagnetic data and identify data points differing sufficiently from the expected background to be considered candidates for a detection.

Finally, measurements of magnetic fields from other planets may be considered. A natural choice is Jupiter~\cite{Davis_1975, Yan_2024}, whose internal magnetic field is $\approx 20$ times stronger on its surface than Earth's~\cite{Connerney}. Although not explicitly stated, the sensitivity~\eqref{eq_estimate} scales with the planetary magnetic moment and a rough -- and optimistic -- estimate of the bounds using Jovian magnetic data would be
\begin{equation} \label{eq_estimate_Jup}
|{\bf k}_J| \sim |{\bf k}_\oplus| \sqrt{ \left( \frac{ R_J }{ R_\oplus } \right) \left( \frac{ \mu_\oplus }{ \mu_J } \right) \left( \frac{ \delta\mathcal{B}_J }{ \delta\mathcal{B}_\oplus } \right) } \, ,
\end{equation}
where $R_J$ and $\mu_J$ are Jupiter's (mean) radius and magnetic moment, respectively, and $\delta\mathcal{B}_J$ is a typical difference between model and data of the Jovian magnetic field. Given that $R_J/R_\oplus \approx 11$ and $\mu_J/\mu_\oplus \approx 2 \times 10^4$~\cite{Connerney, Russell_jup}, we may naively expect our bounds to improve at most by a factor of $\approx 40$ if $\delta\mathcal{B}_J \approx \delta\mathcal{B}_\oplus$, but probably a bit less due to larger uncertainties in both the modelling of the Jovian field and lower quality of the measurements. This, and the other avenues mentioned in the previous paragraph, are worth exploring and will be reported elsewhere.

%%%%%%%%%%%%%%%%%%%%%%%%%%%%%%
\section*{Acknowledgments}
We are grateful to C. Beggan (BGS, Scotland) and C. Finlay (DTU, Denmark) for their kind assistance with the latest geomagnetic data and field models. We also thank M.M. Candido for carefully reading the manuscript. P.C.M., C.A.D.Z. and G.F.C. are indebted
to Marina, Katharina and Karoline Selbach, Lalucha Parizek
and Crispim Augusto, and Nath\'alia de Abreu, respectively, for insightful discussions. The authors acknowledge ESA for the provision of Swarm data. The results presented in this paper rely on data collected 
at magnetic observatories. We thank the national institutes that 
support them and INTERMAGNET for promoting high standards of 
magnetic observatory practice (\url{www.intermagnet.org}). This work was partly funded by the Coordena\c{c}\~{a}o de Aperfei\c{c}oamento de Pessoal de N\'ivel Superior - Brasil (CAPES) - Finance Code 001. C.A.D.Z. is partially supported by Conselho Nacional de Desenvolvimento Cient\'ifico e Tecnol\'ogico (CNPq) under the grant no.~305610/2025-2. C.A.D.Z. is also funded by Funda\c{c}\~{a}o Carlos Chagas Filho de Amparo \`a Pesquisa do Estado do Rio de Janeiro (Faperj) under Grant no.~E-26/201{.}447/2021 (Programa Jovem Cientista do Nosso Estado).

%%%%%%%%%%%%%%%%%%%%%%%%%%%%%%%%%%%%%%%%%%%%
\appendix

%%%%%%%%%%%%%%%%%%%%%%
\section{Components of the CFJ background in the SCF} \label{app_SCF}
\indent

A convenient choice of reference frame is the Sun-centered frame (SCF)~\cite{Kosteleck_2011, Kosteleck__2002, Bluhm_2003}, which is nearly inertial and also known as Sun-centered celestial equatorial frame. Its coordinates $\{T, X,Y,Z \}$ are defined with the $X-Y$-plane parallel to Earth's equatorial plane; the origin of the $T$ coordinate, $T=0$, is defined as the 2000 vernal equinox (20 March 2000, at 07:35~UTC), when Earth crossed the negative $X$-axis. The eccentricity of Earth's orbit is $\approx 0.0167$, so we will approximate it as circular with radius $L_\oplus \approx 1.5 \times 10^8$~km. The $Z$-axis is parallel to Earth's rotational axis $\hat{{\bm \omega}}_\oplus$, which is itself tilted an angle $\eta \approx 23.4^\circ$ relative to its orbital plane. Therefore, denoting $\Psi = \Omega_\oplus T$, the vector describing the translational motion of Earth's center in the SCF is 
\begin{equation} \label{eq_X_earth}
{\bf X}_\oplus = -L_\oplus \left(
\begin{array}{c}
c_\Psi
\\
c_\eta s_\Psi
\\
s_\eta s_\Psi
\end{array}
\right) \, ,
\end{equation}
where $\Omega_\oplus = 2\pi/(365.25 \, d_{\rm sid}) \approx 2.0 \times 10^{-7}$~rad/s is Earth's mean orbital frequency and $d_{\rm sid} = 23.93$h is the sidereal day. Earth's center translates counterclockwise when seen from the positive $Z$-axis with a velocity in the SCF ${\bm \beta}_{\oplus} = d{\bf X}_{\oplus}/dT$ given by
\begin{equation} \label{eq_beta_Earth}
{\bm \beta}_{\oplus} = \beta_\oplus \left(
\begin{array}{c}
s_\Psi 
\\
-c_\eta c_\Psi
\\
-s_\eta c_\Psi
\end{array}
\right) \, 
\end{equation}
with $\beta_{\oplus} = L_\oplus \Omega_\oplus \approx 10^{-4}$.

We wish to obtain the Lorentz transformation connecting the components of the CFJ background as measured in an Earth-bound laboratory or a satellite to the truly constant components in the SCF. To this end we must determine the vector connecting Earth's center to the position of the measurement apparatus, from which we can extract the velocity relative to the SCF, as well as the relative orientation of the axes in the two frames. We start with an observer on the surface and conclude with an apparatus on board of a satellite in low-Earth orbit.

%%%%%%%%%%%%%%%%%%%%%%%%%%%%%
\subsection{Earth-based laboratory} \label{app_lab}
\indent

The standard reference frame has coordinates $\{t,x,y,z\}$ such that the $x$-axis points towards the geographic south ($+\hat{{\bm \theta}}$), the $y$-axis eastward ($+\hat{{\bm \varphi}}$), and the $z$-axis vertically upward ($+\hat{{\bf r}}$). The laboratory rotates with an angular frequency $\omega_\oplus = 2\pi/d_{\rm sid} \approx 7.3 \times 10^{-5}$~rad/s and is connected to Earth's center by the vector 
\begin{equation} \label{eq_X_earth_lab}
{\bf X}_{\oplus, {\rm lab}} = R_\oplus \left(
\begin{array}{c}
s_\theta c_{\psi_\oplus}
\\
s_\theta s_{\psi_\oplus}
\\
c_\theta 
\end{array}
\right) \, ,
\end{equation}
where we kept the shorthand notation introduced earlier: $s_\gamma \equiv \sin\gamma$ and $c_\gamma \equiv \cos\gamma$. The local sidereal time $T_\oplus$ starts when the $Y$- and $y$-axes coincided, differing from $T$ by a constant for each experiment: $T_\oplus = T - T_{\oplus,0}$. As in the main text, we defined $\psi_\oplus = \omega_\oplus T_\oplus$ for convenience. The position of the laboratory in the SCF is given by ${\bf X}_{{\rm lab}} = {\bf X}_\oplus + {\bf X}_{\oplus, {\rm lab}}$. Knowing that $\beta_{L} = R_\oplus \omega_\oplus s_\theta \lesssim 1.5 \times 10^{-6} \ll \beta_\oplus$~\cite{Kosteleck__2002}, its velocity is ${\bm \beta}_{\rm lab} = d{\bf X}_{{\rm lab}}/dT$:
\begin{equation} \label{eq_beta_lab}
{\bm \beta}_{{\rm lab}} = \left(
\begin{array}{c}
\beta_\oplus s_\Psi - \beta_L  s_{\psi_\oplus}
\\
-\beta_\oplus c_\eta c_\Psi + \beta_L  c_{\psi_\oplus}
\\
-\beta_\oplus s_\eta c_\Psi
\end{array}
\right) \, ,
\end{equation}

Let us now consider the relative orientation of the standard laboratory frame $\{ x,y,z \}$ fixed to the surface relative to the SCF. An active (particle) transformation can be performed by assuming that both frames are initially aligned. Take a point $P$ in the North Pole and first rotate it by an angle $\theta$ along the $Y$-axis with $R_Y(\theta)$ and then by an angle $\varphi$ along the $Z$-axis with $R_Z(\varphi)$. This point will now have spherical coordinates $\{\theta, \varphi \}$. In order to express these coordinates in the SCF we take the passive (observer) transformation $\mathcal{R}_{\rm lab} = \left[ R_Z(\varphi) R_Y(\theta) \right]^T$ with~\cite{Kosteleck__2002}
\begin{equation} \label{eq_rot_matrix}
\mathcal{R}_{\rm lab} =\left(
\begin{array}{ccc}
c_\theta c_{\psi_\oplus}
& 
c_\theta s_{\psi_\oplus}
& 
-s_\theta
\\
-s_{\psi_\oplus} 
& 
c_{\psi_\oplus} 
& 
0
\\
s_\theta c_{\psi_\oplus}
&
s_\theta s_{\psi_\oplus}
& 
c_\theta
\end{array}
\right) \, .
\end{equation}

The Lorentz transformation $\Lambda$ between the SCF and the standard laboratory frame may be explicitly factored as the combination of a transformation including only spatial rotations, $\Lambda_R$, and another only with boosts, $\Lambda_\beta$: $\Lambda = \Lambda_R \Lambda_\beta$. We then have
\begin{equation} \label{eq_lambdas_lab}
\left(\Lambda_R \right)^\mu_{\,\,\,\nu} = \left(
\begin{array}{cc}
1 & 0 \\
0 & \mathcal{R}_{\rm lab} 
\end{array}
\right) \,\,\, {\rm and} \,\, \,\left(\Lambda_\beta \right)^\mu_{\,\,\,\nu} \approx \left(
\begin{array}{cc}
1 & -{\bm \beta}_{\rm lab}^T \\
-{\bm \beta}_{\rm lab} & 1
\end{array}
\right) \, ,
\end{equation}
where the entries ``1" represent identity matrices of adequate dimension. All velocities involved are non relativistic, $\beta_{\rm lab} \ll 1$, therefore allowing us to set $\gamma = \sqrt{1 - \beta_{\rm lab}^2} \approx 1$. We shall thus limit our analysis to first-order effects in the velocities. Denoting the elements of the full Lorentz transformation matrix by $\Lambda^{\mu}_{\,\,\nu}$ with $\mu = \{t,x,y,z\}$ and $\nu = \{T,X,Y,Z\}$ for the standard laboratory frame fixed to the terrestrial surface and the components of in the SCF, we have $\Lambda^{0}_{\,\,T} = 1$, $\Lambda^{0}_{\,\,J} = -{\bm \beta}_{\rm lab}^{J}$, $\Lambda^{i}_{\,\,T} = - (\mathcal{R}_{\rm lab}\,{\bm \beta}_{\rm lab})^{i}$ and $\Lambda^{i}_{\,\,J} = (\mathcal{R}_{\rm lab})^{i}_{\,\,J}$~\cite{Kosteleck__2002}. The components of the CFJ background in the standard laboratory frame may be thus written as
\begin{eqnarray} 
k^0 & \approx & k_{\rm SCF}^{\rm T} - {\bm \beta}_{\rm lab} \cdot {\bf k}_{\rm SCF} \, , \label{k0_SCF_app} \\
{\bf k}^i & \approx & - (\mathcal{R}_{\rm lab} \, {\bm \beta}_{\rm lab})^i k_{\rm SCF}^{\rm T} +  (\mathcal{R}_{\rm lab} \, {\bf k}_{\rm SCF})^i  \, .\label{ki_SCF_app}
\end{eqnarray}

Imposing that $k_{\rm SCF}^{\rm T} \equiv 0$ and $\beta_L \ll \beta_\oplus$, the time component in the laboratory frame becomes
\begin{equation} \label{eq_app_k0_lab}
k_0 \approx -\beta_\oplus \left[ k_X s_\Psi - \left( c_\eta k_Y + s_\eta k_Z \right) c_\Psi \right] \, .
\end{equation}
Simply writing ${\bf k}^i = k_i$ for convenience, the explicit expressions for the spatial components in the standard Earth-bound frame are then
\begin{eqnarray}
k_x & = & k_X c_\theta c_{\psi_\oplus} + k_Y c_\theta s_{\psi_\oplus} - k_Z s_\theta  \, , \label{eq_app_kx_lab} \\
k_y & = & -k_X s_{\psi_\oplus} + k_Y c_{\psi_\oplus} \, , \label{eq_app_ky_lab} \\
k_z & = & k_X s_\theta c_{\psi_\oplus} + k_Y s_\theta s_{\psi_\oplus} + k_Z c_\theta  \, . \label{eq_app_kz_lab} 
\end{eqnarray}
Comparing $k_0$ and $k_i$ it is clear that $k_0$ is significantly suppressed by $\beta_\oplus$ and oscillates much more slowly.

%%%%%%%%%%%%%%%%%%%%%%%%%%%%%
\subsection{Satellite-based laboratory} \label{app_sat}
\indent

The case of a satellite is analogous to that of an Earth-bound laboratory, but here the satellite's orbit does not have a constant polar angle (analogous, but not equal to, geographic co-latitude) as measured from a non-rotating frame fixed at Earth's center. The first step towards the Lorentz transformation connecting the SCF coordinates $\{ T, X, Y, Z \}$ to those of a standard satellite-borne laboratory $\{ t,x,y,z \}$ is to determine the vector from the origin of the SCF to the center of the satellite.

Following Ref.~\cite{Kosteleck__2002}, we define the coordinates in the standard satellite frame $\{ t,x,y,z \}$ so that the $z$-axis is along the satellite velocity ${\bm \beta}_{\oplus, s}$ relative to Earth's center, the $x$-axis points towards Earth's center and the $y$-axis completes the right-handed coordinate system. Here $\beta_{\oplus, s} = |{\bm \beta}_{\oplus, s}| = r_s \omega_s$, where $r_s$ is the mean orbital radius $r_s$ and $\omega_s$ is the satellite's orbital angular frequency around Earth. The orbit, assumed circular, is further characterized by two angles: $\zeta$, the angle between the orbital and Earth’s equatorial planes, and $\alpha$, the azimuthal angle where the satellite's orbit crosses the equatorial $X-Y$-plane. This happens twice per orbit, so $\alpha$ is defined as the angle measured from the $X$-axis when the satellite is in an ascending trajectory (ascending node).

Let us start by considering the orbit as lying parallel to the $X-Y$-plane with $x_o-y_o$-axes aligned to those of the SCF. At an initial time $T_{s,0}$ the satellite is at $x_o(T_{s,0}) = r_s$ and $y_o(T_{s,0}) = 0$ moving counterclockwise when seen from above, so that $x_o(T_s) = r_s \cos\psi_s$ and $y_o(T_s) = r_s\sin\psi_s$ with $\psi_s = \omega_s T_s$ and $T_s = T - T_{s,0}$. To reach the orientation of the orbit shown in Fig.~2 of Ref.~\cite{Bluhm_2003}, the position of the satellite relative to Earth's center must be first rotated by $\zeta$ around the $X$-axis -- this leaves the position of the satellite at $T_{s,0}$ on the positive $X$-axis, but removes the orbital plane from the $X-Y$-plane. Therefore, $\zeta$ is the inclination of the orbit relative to the equatorial plane. Next, we apply a rotation of an angle $\alpha$ around the $Z$-axis to place the intersection of the satellite's and Earth's orbital planes in the correct position, yielding~\cite{Bluhm_2003}
\begin{equation} \label{eq_X_earth_sat}
{\bf X}_{\oplus, {\rm s}} = r_s \left(
\begin{array}{c}
c_\alpha c_{\psi_s} - c_\zeta s_\alpha s_{\psi_s}
\\
s_\alpha c_{\psi_s} + c_\zeta c_\alpha s_{\psi_s}
\\
s_\zeta s_{\psi_s}
\end{array}
\right) \, .
\end{equation}

The azimuthal angle $\alpha$ drifts at a rate $\omega_{\alpha}$, cf. Eq.~\eqref{eq_nodal_drift_rate}, so $\alpha = \alpha_0 + \omega_\alpha T_s$. At $T = T_{s,0}$ the satellite is on the $X-Y$ plane at $X_{\oplus, {\rm s}} = r_s c_\alpha$ and $Y_{\oplus, {\rm s}} = r_s s_\alpha$. Therefore, $T_{s,0}$ may be freely chosen as any moment when the satellite crosses the equatorial plane in its ascending path, whereupon the satellite crosses the ascending node at $\alpha = \alpha_0$. Since it is impossible to determine $\alpha_0$ without a detailed reconstruction of the satellite's orbit in the SCF, we must leave it as an unknown in $\alpha = \alpha_0 + \omega_{\alpha}T_s$ and allow it to vary within $\left[ 0, 2\pi \right]$.

The position of the center of the satellite relative to the origin of the SCF is ${\bf X}_{s} = {\bf X}_\oplus + {\bf X}_{\oplus, s}$ and its velocity vector, ${\bm \beta}_{s} = d{\bf X}_{s}/dT$, is
\begin{eqnarray} \label{eq_beta_sat}
{\bm \beta}_{s} & = & \left(
\begin{array}{c}
 \beta_\oplus s_\Psi - \beta_{\oplus, s} (c_\alpha s_{\psi_s} + c_\zeta s_\alpha c_{\psi_s})
\\
-\beta_\oplus c_\eta c_\Psi - \beta_{\oplus, s} (s_\alpha s_{\psi_s} - c_\zeta c_\alpha c_{\psi_s})
\\
-\beta_\oplus s_\eta c_\Psi + \beta_{\oplus, s} s_\zeta c_{\psi_s}
\end{array}
\right) +
\beta_\alpha \left(
\begin{array}{c}
-s_\alpha c_{\psi_s} - c_\zeta c_\alpha s_{\psi_s}
\\
c_\alpha c_{\psi_s} - c_\zeta s_\alpha s_{\psi_s}
\\
0
\end{array}
\right) \, ,
\end{eqnarray}
where $\beta_{\oplus, s} = r_s \omega_s$ and $\beta_\alpha = r_s \omega_\alpha$; the results from Refs.~\cite{Bluhm_2003, Kosteleck__2002} are recovered for $\omega_\alpha = 0$. We have $\omega_\alpha \ll \omega_s$, that is, $\beta_{\oplus, s} \gg \beta_\alpha$ and the second term of Eq.~\eqref{eq_beta_sat} may be neglected.

In the standard satellite frame the $x$-axis points toward Earth's center, $\hat{{\bf x}} = - {\bf X}_{\oplus, {\rm s}}/r_s$, and the $z$-axis is along the velocity of the satellite relative to Earth's center, $\hat{{\bf z}} = \hat{{\bm \beta}}_{\oplus, s} = {\bm \beta}_{\oplus, s}/\beta_{\oplus, s}$. The $y$-axis points along the orbital angular momentum vector, so that $\hat{{\bf y}} = \hat{{\bm \omega}}_s = \hat{{\bf z}} \times \hat{{\bf x}}$:
\begin{equation} \label{eq_y_sat}
\hat{{\bf y}} = \left(
\begin{array}{c}
s_\alpha s_\zeta
\\
-c_\alpha s_\zeta
\\
c_\zeta 
\end{array}
\right) \, ,
\end{equation}
which, in the absence of nodal precession, is time independent, as it should. The passive (observer) transformation $\mathcal{R}$ expressing the components of the CFJ background in the standard satellite frame in terms of those in the SCF is given by~\cite{Kosteleck__2002}
\begin{equation} \label{eq_rot_matrix_sat}
\mathcal{R}_s \! = \! \left( \!\!
\begin{array}{ccc}
-c_\alpha c_{\psi_s} + c_\zeta s_\alpha s_{\psi_s} 
& 
-s_\alpha c_{\psi_s} - c_\zeta c_\alpha s_{\psi_s}
& 
-s_\zeta s_{\psi_s}
\\
s_\alpha s_\zeta
& 
-c_\alpha s_\zeta
& 
c_\zeta 
\\
-c_\alpha s_{\psi_s} - c_\zeta s_\alpha c_{\psi_s} 
& 
-s_\alpha s_{\psi_s} + c_\zeta c_\alpha c_{\psi_s} 
& 
s_\zeta c_{\psi_s}
\end{array}
\!\! \right).
\end{equation}

The arguments around Eq.~\eqref{eq_lambdas_lab} are valid here and the components of the CFJ background in the satellite frame may be related to those in the SCF by equations analogous to Eqs.~\eqref{k0_SCF_app} and~\eqref{ki_SCF_app}, but using ${\bm \beta}_{s}$ and $\mathcal{R}_s$ instead of ${\bm \beta}_{\rm lab}$ and $\mathcal{R}_{\rm lab}$. An experimental device in the satellite may have a different orientation relative to the standard satellite frame, but throughout this work we assume that the instruments onboard of the satellite are aligned in accordance with the standard satellite reference frame $\{ x,y,z \}$ defined above.

\begin{figure}[t!]
\centering
\begin{minipage}[b]{0.80\linewidth}
\includegraphics[width=\textwidth]{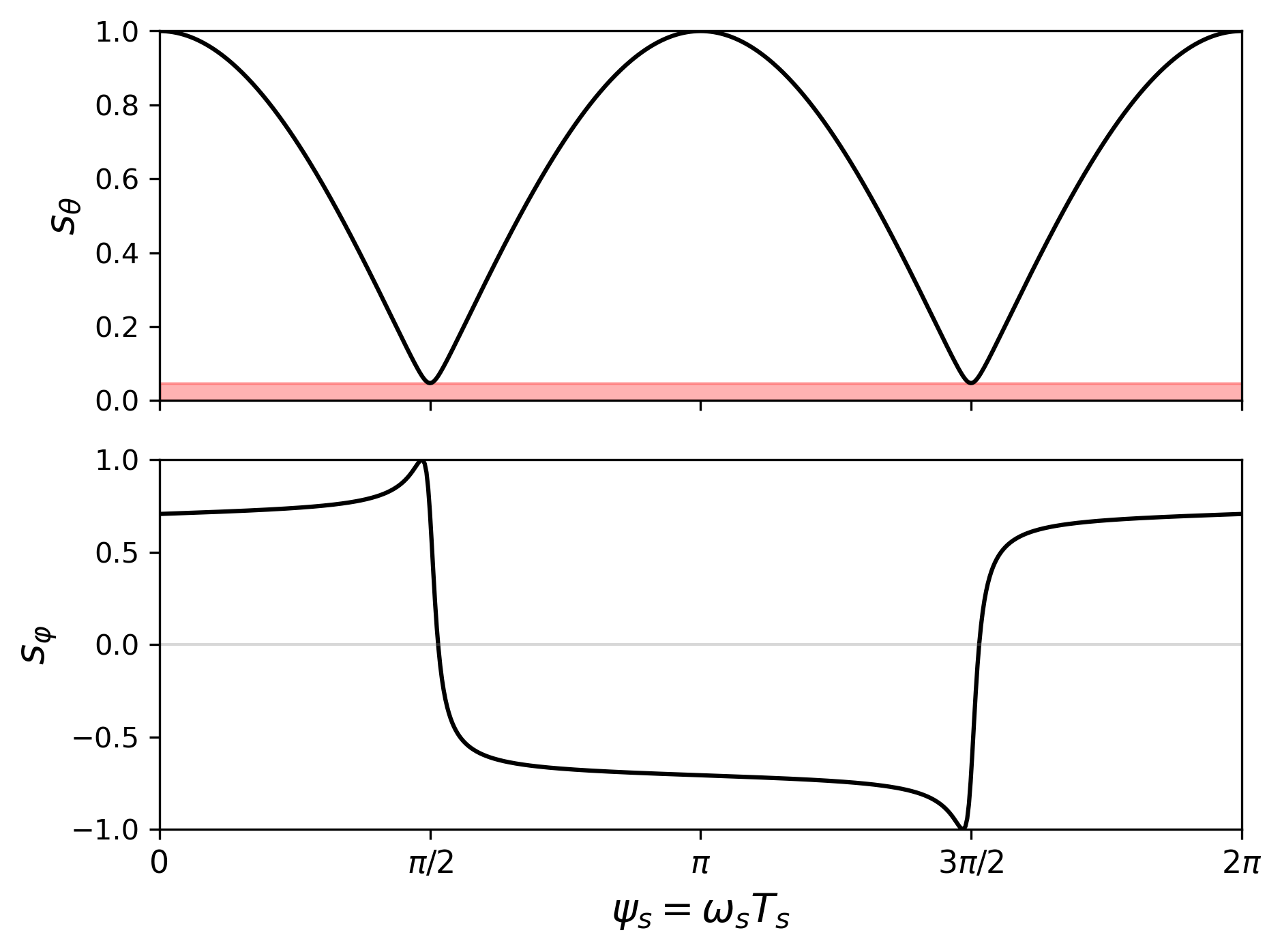}
\end{minipage} \hfill
\caption{Spherical coordinates of an orbit similar to that of Swarm A with inclination $\zeta = 87.3^\circ$, as given by Eqs.~\eqref{eq_app_sin_phi} and~\eqref{eq_app_sine_theta}, here shown as functions of $\psi_s = \omega_s T_s$ for one full revolution starting at the equatorial plane at $\varphi = \alpha = 45^\circ$ and initially ascending towards the North Pole. Note that $s_\theta$ does not reach zero because $\zeta \neq 90^\circ$; the ``cap" that is not covered by the satellite is highlighted in red on the upper panel.  }
\label{fig_sin_cos_Ts}
\end{figure}

The data we use, as well as geomagnetic models describing them, are expressed in Earth-centered spherical coordinates $\{ r,\theta,\varphi\}$, which are in general not aligned with $\{ x,y,z \}$. The $x$-axis of the satellite frame points toward Earth, so that $\hat{{\bf x}} = -\hat{{\bf r}}$, and the remaining unitary vectors in the satellite frame, $\{ \hat{{\bf y}}, \hat{{\bf z}}\}$, are tangential to the sphere with origin in Earth's center. Therefore, these may be expressed as combinations of the standard tangential vectors of spherical coordinates, $\{ \hat{{\bm \theta}}, \hat{{\bm \varphi}}\}$. Let us again consider the non-rotating Earth-centered frame with axes $X-Y-Z$ conveniently aligned with those of the SCF. From the components of ${\bf X}_{\oplus, {\rm s}}$, cf. Eq.~\eqref{eq_X_earth_sat}, the polar and azimuthal angles (measured from the positive $X$-axis) of the satellite at a moment $T_s$ are given by
\begin{eqnarray}
c_\theta & = & s_\zeta s_{\psi_s} \, , \label{eq_app_cos_theta} \\
s_\varphi & = & (s_\alpha c_{\psi_s} + c_\zeta c_\alpha s_{\psi_s})/s_\theta \, , \label{eq_app_sin_phi} \\
c_\varphi & = & (c_\alpha c_{\psi_s} - c_\zeta s_\alpha s_{\psi_s})/s_\theta \, \label{eq_app_cos_phi}
\end{eqnarray}
showing that the largest co-latitude $\theta$ covered by the satellite is determined by its orbital inclination $\zeta$. This explains why satellites are launched with high inclinations to maximize spatial coverage. Since, by definition of the co-latitude, $0 \leq \theta \leq \pi$, we have $0 \leq s_\theta \leq 1$ and, when expressing the sine of the co-latitude from Eq.~\eqref{eq_app_cos_theta}, we may take the positive root: 
\begin{equation} \label{eq_app_sine_theta}
s_\theta = +\sqrt{1 - s_\zeta^2 s_{\psi_s}^2} \, .
\end{equation}
In Fig.~\ref{fig_sin_cos_Ts} we show the time dependence of $s_\theta$ and $s_\varphi$ for $\zeta = 87.3^\circ$, the same as that of Swarm A, and fixed $\alpha$. The spherical components of the CFJ background may thus be concisely expressed in terms of the polar and azimuthal angles as
\begin{eqnarray}
k_r & = & k_X s_\theta c_\varphi + k_Y s_\theta s_\varphi + k_Z c_\theta  \, , \label{eq_app_kr_sat} \\
k_\theta & = & k_X c_\theta c_\varphi + k_Y c_\theta s_\varphi - k_Z s_\theta \, , \label{eq_app_ktheta_sat} \\
k_\varphi & = & -k_X s_\varphi + k_Y c_\varphi  \, . \label{eq_app_kphi_sat} 
\end{eqnarray}

Finally, consider the time component of the CFJ 4-vector background in the standard satellite frame, $k^0$. It is given by 
\begin{equation} \label{eq_app_k0_sat}
k^0 \approx - {\bm \beta}_{\rm s} \cdot {\bf k}_{\rm SCF} \, ,
\end{equation}
analogous to the case of a ground observatory, cf. Eq.~\eqref{k0_SCF_app}, with $k_{\rm SCF}^{\rm T} \equiv 0$. The boost vector, ${\bm \beta}_{\rm s}$, depends on three boost factors, $\beta_{\oplus}$, $\beta_{\oplus ,s}$ and $\beta_\alpha$, cf. Eq.~\eqref{eq_beta_sat}. From Eq.~\eqref{eq_beta_Earth} we have $\beta_{\oplus} = L_\oplus \Omega_\oplus \approx 10^{-4}$, whereas the orbital velocity of the Swarm satellites relative to Earth's center is $\beta_{\oplus ,s} = \omega_s (R_\oplus + h) \approx 2.6 \times 10^{-5}$ and $|\omega_{\alpha}| \approx 7.4 \times 10^{-8}$~rad/s (Swarm~A), implying that $\beta_\alpha = \omega_\alpha r_s \approx 10^{-9}$. These values are arranged according to the hierarchy $\beta_{\oplus} \gtrsim \beta_{\oplus ,s} \gg \beta_\alpha$ and we may keep only the boost contributions depending on $\beta_{\oplus}$ in ${\bm \beta}_{\rm s}$, ending up with the same result as for an Earth-bound laboratory, namely, Eq.~\eqref{eq_app_k0_lab}. Therefore, the time component depends on the orbital parameters only via the orbital inclination $\zeta$ and $\psi_s = \omega_s T_s$, but not via the azimuthal angle of the ascending node $\alpha$.

\begin{figure}[t!]
\centering
\begin{minipage}[b]{0.95\linewidth}
\includegraphics[width=\textwidth]{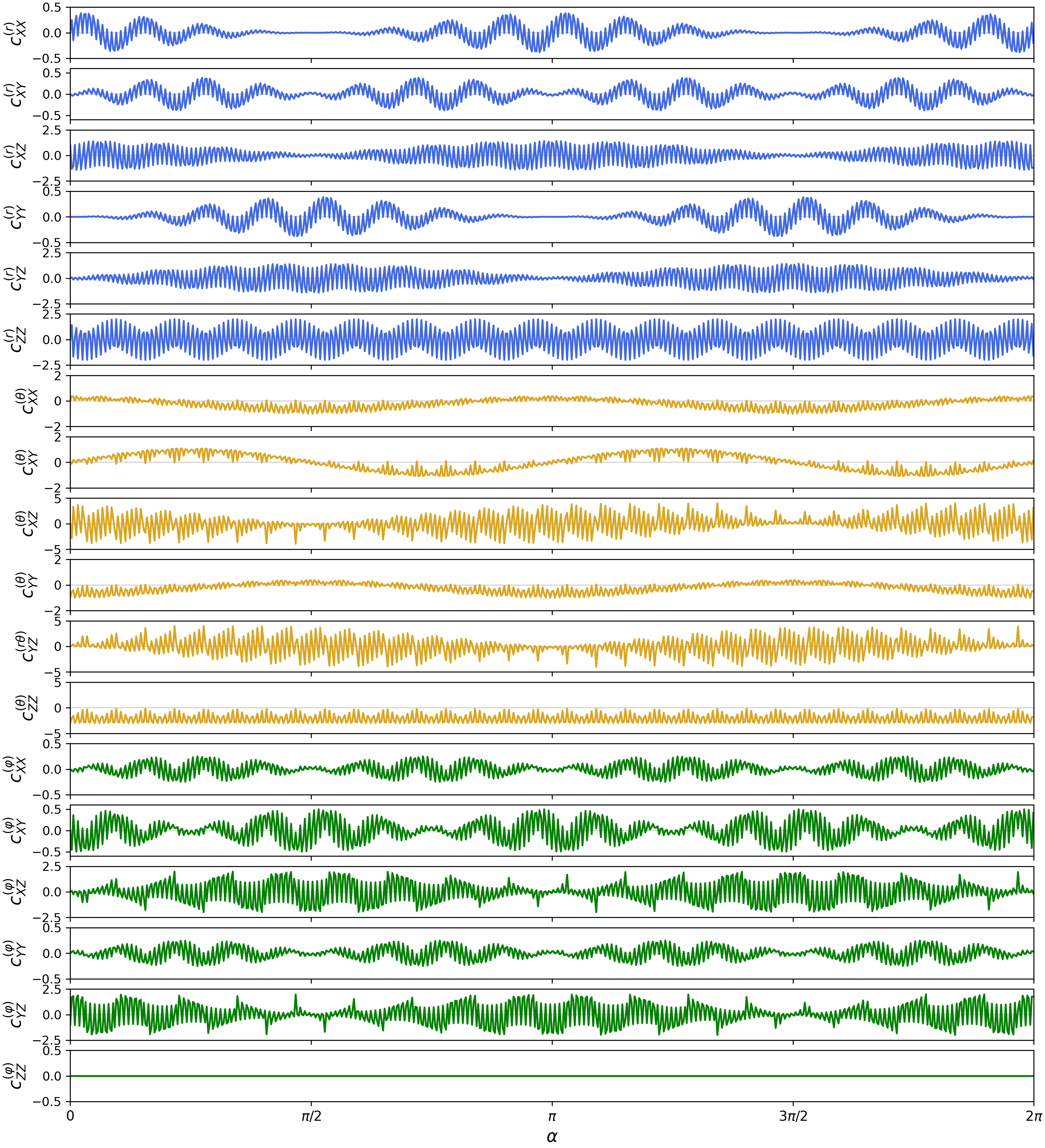}
\end{minipage} \hfill
\caption{Time-dependent coefficients $c^{(u)}_{ab}$ for $u = \{ r, \theta, \varphi \}$ as functions of the ascending node angle $\alpha = \alpha_0 + \omega_\alpha T_s$ over one full period $\tau_\alpha$; for Swarm A this is equivalent to $\approx 2.7$~yr. We take $\alpha$ as free variable, so that a variation of $\alpha$ from zero to $2\pi$ is equivalent to $\approx 2.7$~yr and $\psi_s = (\omega_s/\omega_\alpha)\alpha -(\omega_s/\omega_\alpha)\alpha_0$. The fast oscillations have the period of a full revolution of the satellite, $\tau_s \ll \tau_\alpha$; for Swarm A we have $\tau_s = 94$~min. The combination of the two frequencies give rise to beatings which, for several of the $c^{(u)}_{ab}$, have periods of $\tau_\alpha/2$ (for example, $c^{(r)}_{XX}$, $c^{(\theta)}_{XZ}$ or $c^{(\varphi)}_{YZ}$). Others, such as $c^{(r)}_{ZZ}$, $c^{(\theta)}_{ZZ}$ and $c^{(\varphi)}_{YY}$ display even shorter periods. In any case, these coefficients are approximately periodic with a period of $\mathcal{O}(\tau_\alpha)$.}
\label{fig_c_ij_all}
\end{figure}

With these results we obtain relatively cumbersome expressions, which were compactly written in terms of time-dependent coefficients $c^{(u)}_{ab}$ with $u = \{ r, \theta, \varphi \}$ in Eqs.~\eqref{eq_B_rad_sat} and~\eqref{eq_B_theta_phi_sat}. The coefficients for the radial component are
\begin{eqnarray}
c^{(r)}_{XX} & = & s_\zeta s_{\psi_s} (c_\alpha c_{\psi_s} - s_\alpha c_\zeta s_{\psi_s})^2  \, , \label{eq_app_c_XX_rad} \\ 
c^{(r)}_{XY} & = &  2s_\zeta s_{\psi_s} (s_\alpha c_{\psi_s} + c_\alpha c_\zeta s_{\psi_s})(c_\alpha c_{\psi_s} - s_\alpha c_\zeta s_{\psi_s}) \, ,  \\ 
c^{(r)}_{XZ} & = & (1 + 2s_\zeta^2 s_{\psi_s}^2) (c_\alpha c_{\psi_s} - s_\alpha c_\zeta s_{\psi_s}) \, ,  \\ 
c^{(r)}_{YY} & = &  s_\zeta s_{\psi_s} (s_\alpha c_{\psi_s} + c_\alpha c_\zeta s_{\psi_s})^2  \, ,  \\ 
c^{(r)}_{YZ} & = &  (1 + 2s_\zeta^2 s_{\psi_s}^2) (s_\alpha c_{\psi_s} + c_\alpha c_\zeta s_{\psi_s}) \, ,  \\ 
c^{(r)}_{ZZ} & = &  s_\zeta s_{\psi_s}(1 + s_\zeta^2 s_{\psi_s}^2) \, , \label{eq_app_c_ZZ_rad}  \\ 
s_\theta \, c^{(\theta)}_{XX} & = &  -\left[ c_{\psi_s} (s_\alpha + c_\alpha s_\zeta s_{\psi_s}) + c_\zeta s_{\psi_s} (c_\alpha - s_\alpha s_\zeta s_{\psi_s}) \right] \nonumber \\
& \times & \left[ c_{\psi_s} (s_\alpha - c_\alpha s_\zeta s_{\psi_s}) + c_\zeta s_{\psi_s} (c_\alpha + s_\alpha s_\zeta s_{\psi_s}) \right]  \, , \label{eq_app_c_XX_theta} \\ 
s_\theta \,  c^{(\theta)}_{XY} & = & 2(1 + s_\zeta^2 s_{\psi_s}^2) (s_\alpha c_{\psi_s} + c_\alpha c_\zeta s_{\psi_s})(c_\alpha c_{\psi_s} - s_\alpha c_\zeta s_{\psi_s}) \, ,   \\ 
s_\theta \,  c^{(\theta)}_{XZ} & = & 2s_\zeta s_{\psi_s}  (1 + s_\zeta^2 s_{\psi_s}^2) (c_\alpha c_{\psi_s} - s_\alpha c_\zeta s_{\psi_s})  \, , \\ 
s_\theta \,  c^{(\theta)}_{YY} & = &  \frac{1}{2}c_\zeta s_{2\psi_s} s_{2\alpha} + \frac{1}{2}s_{\psi_s}^2 \left[ s_{2\alpha} s_{2\psi_s}c_\zeta s_\zeta^2 - 2s_\alpha^2 (c_\zeta^2 - s_\zeta^2 c_{\psi_s}^2) \right] - c_\alpha^2 (c_{\psi_s}^2 - s_\zeta^2 c_\zeta^2s_{\psi_s}^4)   \, , \\ 
s_\theta \,  c^{(\theta)}_{YZ} & = & 2s_\zeta s_{\psi_s}  (1 + s_\zeta^2 s_{\psi_s}^2) (s_\alpha c_{\psi_s} + c_\alpha c_\zeta s_{\psi_s}) \, ,  \\ 
s_\theta \,  c^{(\theta)}_{ZZ} & = & -3 + s_\zeta^2 s_{\psi_s}^2 (2 + s_\zeta^2 s_{\psi_s}^2) \, \label{eq_app_c_ZZ_theta} \, ,  \\
s_\theta \,  c^{(\varphi)}_{XX} & = & s_\zeta s_{\psi_s} (s_\alpha c_{\psi_s} + c_\alpha c_\zeta s_{\psi_s})(c_\alpha c_{\psi_s} - s_\alpha c_\zeta s_{\psi_s})   \, , \label{eq_app_c_XX_phi}  \\ 
s_\theta \,  c^{(\varphi)}_{XY} & = & s_\zeta s_{\psi_s} \left[ s_{2\alpha} c_\zeta s_{2\psi_s} - c_{2\alpha} (c_{\psi_s}^2 - c_\zeta^2 s_{\psi_s}^2)  \right] \, ,   \\ 
s_\theta \,  c^{(\varphi)}_{XZ} & = & (1 + s_\zeta^2 s_{\psi_s}^2) (s_\alpha c_{\psi_s} + c_\alpha c_\zeta s_{\psi_s}) \, ,   \\ 
s_\theta \,  c^{(\varphi)}_{YY} & = & -s_\zeta s_{\psi_s} (s_\alpha c_{\psi_s} + c_\alpha c_\zeta s_{\psi_s})(c_\alpha c_{\psi_s} - s_\alpha c_\zeta s_{\psi_s}) \, ,   \\ 
s_\theta \,  c^{(\varphi)}_{YZ} & = & -(1 + s_\zeta^2 s_{\psi_s}^2) (c_\alpha c_{\psi_s} - s_\alpha c_\zeta s_{\psi_s}) \, ,   \\ 
s_\theta \,  c^{(\varphi)}_{ZZ} & = &  0 \, . \label{eq_app_c_ZZ_phi}
\end{eqnarray}
Note that $c^{(r)}_{XZ}$ and $c^{(r)}_{YZ}$ differ essentially by the shift $\alpha \rightarrow \alpha + \pi/2$; the same is valid for $c^{(\varphi)}_{XZ}$ and $c^{(\varphi)}_{YZ}$. This explains the shifts between the red and green curves on the upper and lower panels of Fig.~\ref{fig_b_alpha_0}. All these coefficients are shown in Fig.~\ref{fig_c_ij_all}.

%%%%%%%%%%%%%%%%%%%%%%
\section{The 4-potential of a point-like magnetic dipole} \label{app_details}
\indent

In Sec.~\ref{sec_dipole} we discussed the electromagnetic fields produced by a static magnetic dipole ${\bm \mu}$ located at ${\bf x}_0$. Some necessary mathematical results are supplied in App.~\ref{sec_app_maths}, whereas in App.~\ref{sec_app_pots} we obtain expressions that allow us to calculate the potentials given the propagator.

%%%%%%
\subsection{Useful formulae}\label{sec_app_maths}

The source is located at ${\bf x}_0$, here taken as the origin, and the field is evaluated at ${\bf x}$. The position vector is ${\bf r} = {\bf x} - {\bf x}_0$, $r = |{\bf r}|$ and $\hat{{\bf r}} = {\bf r}/r$, which satisfy a few relevant identities:
\begin{eqnarray}
{\bm \nabla} r & = & \hat{{\bf r}} \, , \quad  {\bm \nabla} r^{-1} = -\hat{{\bf r}}/r^2  \, , \\
\,{\bm \nabla} \cdot \hat{{\bf r}} & = & 2/r  \,  , \quad {\bm \nabla}_a \hat{{\bf r}}_b  = \frac{1}{r} \left( \delta_{ab} - \hat{{\bf r}}_a\hat{{\bf r}}_b \right) \, .
\end{eqnarray}
Moreover, the following Fourier integrals are useful:
\begin{eqnarray}
I_2({\bf a}) & = & \int \frac{d^3 {\bf p}}{(2\pi)^3} \frac{e^{i {\bf p}\cdot{\bf a}}}{ {\bf p}^2 } = \frac{1}{4\pi |{\bf a}|} \label{eq_int_2} \, , \\
I_4({\bf a}) & = & \int \frac{d^3 {\bf p}}{(2\pi)^3} \frac{e^{i {\bf p}\cdot{\bf a}}}{ {\bf p}^4 } = -\frac{ |{\bf a}|}{8\pi} \label{eq_int_4} \, , \\
I_6({\bf a}) & = & \int \frac{d^3 {\bf p}}{(2\pi)^3} \frac{e^{i {\bf p}\cdot{\bf a}}}{ {\bf p}^6 } = \frac{ |{\bf a}|^3 }{96\pi} \label{eq_int_6} \, ,
\end{eqnarray}
from which other, related integrals may be derived via differentiation with respect to the components of ${\bf a}$:
\begin{eqnarray}
\int \frac{d^3 {\bf p}}{(2\pi)^3} \frac{ {\bf p}_i {\bf p}_j e^{i {\bf p}\cdot{\bf a}}}{ {\bf p}^2 } & = & \frac{1}{4\pi |{\bf a}|^3} \left( \delta_{ij} - 3\hat{{\bf a}}_i \hat{{\bf a}}_j \right), \label{eq_int_2_pp} \\
\int \frac{d^3 {\bf p}}{(2\pi)^3} \frac{ {\bf p}_i {\bf p}_j e^{i {\bf p}\cdot{\bf a}}}{ {\bf p}^4 } & = & \frac{1}{8\pi |{\bf a}|} \left( \delta_{ij} - \hat{{\bf a}}_i \hat{{\bf a}}_j \right)  \label{eq_int_4_pp} \, , \\
\int \frac{d^3 {\bf p}}{(2\pi)^3} \frac{ {\bf p}_i {\bf p}_j e^{i {\bf p}\cdot{\bf a}}}{ {\bf p}^6 } & = & -\frac{|{\bf a}|}{32\pi} \left( \delta_{ij} + \hat{{\bf a}}_i \hat{{\bf a}}_j \right)  \label{eq_int_6_pp2} \, , \\
\int \frac{d^3 {\bf p}}{(2\pi)^3} \frac{ {\bf p}_i {\bf p}_j {\bf p}_n {\bf p}_m  e^{i {\bf p}\cdot{\bf a}}}{ {\bf p}^6 } & = & \frac{ 1 }{32\pi |{\bf a}| } \left[  h_{ij}(\hat{{\bf a}}) h_{nm}(\hat{{\bf a}})  + h_{in}(\hat{{\bf a}}) h_{jm}(\hat{{\bf a}})  + h_{im}(\hat{{\bf a}}) h_{nj}(\hat{{\bf a}})  \right]  \label{eq_int_6_pp} \, 
\end{eqnarray}
with $h_{ij}(\hat{{\bf a}}) = \delta_{ij} - \hat{{\bf a}}_i \hat{{\bf a}}_j$; note the symmetry under permutation of the indices.

%%%%
\subsection{Scalar and vector potentials}\label{sec_app_pots}
\indent

The 4-potential can be obtained by solving Eq.~\eqref{eq_wave_eq_p} using the momentum-space propagator~\eqref{eq_D_inv}. For the scalar potential we need $D^{-1}_{0i}({\bf p})$ and the position-space propagator becomes
\begin{equation} \label{eq_app_Gi}
{\bf G}_i({\bf x} - {\bf x}') = \int \frac{d^3 {\bf p}}{(2\pi)^3} D^{-1}_{0i}({\bf p}) e^{i {\bf p}\cdot({\bf x} - {\bf x}')} \, 
\end{equation}
with
\begin{equation}\label{eq_D_0i}
D^{-1}_{0i}({\bf p}) = -2i\epsilon_{iab} \frac{{\bf k}_a {\bf p}_b}{{\bf p}^4} - \frac{4k_0}{{\bf p}^4} \! \left( {\bf k}_i - {\bf k}_j \frac{{\bf p}_i{\bf p}_j}{{\bf p}^2} \right) \, .
\end{equation}
The scalar analog to Eq.~\eqref{eq_Ai} is 
\begin{equation}  \label{eq_app_A0_1}
A_0({\bf x}) = -\epsilon_{ijk} \int d^3 {\bf x}' \left[ {\bm \nabla}'_j {\bf M}_k({\bf x}') \right] {\bf G}_i({\bf x} - {\bf x}') \, ,
\end{equation}
where ${\bf J}({\bf x}') = {\bm \nabla}' \times {\bf M}({\bf x}')$; ${\bm \nabla}'$ acts on primed coordinates. Using $\partial (f g) = f (\partial g) + g (\partial f)$, Eq.~\eqref{eq_app_A0_1} becomes
\begin{equation}  \label{eq_app_A01}
A_0({\bf x}) = \int d^3 {\bf x}' \left[ {\bm \nabla}' \cdot \left( {\bf G} \times {\bf M} \right)  - {\bf M} \cdot \left( {\bm \nabla}' \times {\bf G} \right)  \right] \, .
\end{equation}
The first term may be converted into a surface integral via Gauss' theorem, but it is zero because ${\bf M} = 0$ everywhere except at ${\bf x}_0$. Plugging ${\bf M}({\bf x}') = {\bm \mu} \delta^3 ({\bf x}' - {\bf x}_0)$ into the second term, noting that only the combination ${\bf x} - {\bf x}'$ appears and using ${\bm \nabla}' = - {\bm \nabla}$, we get
\begin{equation}  \label{eq_app_A02}
A_0({\bf x}) = {\bm \mu} \cdot \left[ {\bm \nabla} \times {\bf G}({\bf x} - {\bf x}_0) \right] \, .
\end{equation}

The vector potential can be calculated from Eq.~\eqref{eq_Ai}, so let us define ${\bf F}_{(i)}({\bf x} - {\bf x}')$ whose $j$-th component is ($i$ fixed)
\begin{equation} \label{eq_app_Fij}
{\bf F}_{(i)j}({\bf x} - {\bf x}') = \int \frac{d^3 {\bf p}}{(2\pi)^3} D^{-1}_{ij}({\bf p}) e^{i {\bf p}\cdot({\bf x} - {\bf x}')}  \, .
\end{equation}
Following the same steps as done for the scalar potential we finally find 
\begin{equation}  \label{eq_app_Ai2}
{\bf A}_i({\bf x}) = -{\bm \mu} \cdot \left[ {\bm \nabla} \times {\bf F}_{(i)}({\bf x} - {\bf x}_0) \right] \, .
\end{equation}
%which is most easily evaluated by expressing its right-hand side in component form.

%%%%%%%%%%%%%%%%%%%%%%
\section{Electric field} \label{app_E_field}
\indent 

Our focus on the main text was the magnetic field, but in the CFJ electrodynamics a static magnetic dipole will also originate an electric field. Starting with Eq.~\eqref{eq_app_A02} and using the formulae from App.~\ref{sec_app_maths} we get
\begin{equation}
A_0({\bf x}) = \frac{1}{4\pi r} \left[ ( {\bf k}\cdot{\bm \mu} ) + ( {\bf k}\cdot\hat{{\bf r}} ) ( {\bm \mu}\cdot\hat{{\bf r}} )  \right] + \frac{k_0}{2\pi} {\bm \mu}\cdot\left( \hat{{\bf r}} \times {\bf k} \right) \label{eq_A0_final} \, .
\end{equation}
In Maxwell's electrodynamics a static magnetic dipole does not produce an electric field -- the scalar potential is a constant, usually taken to be zero. As shown above, this is not the case here. Importantly, $A_0$ is linear in ${\bf k}$ with a second-order correction involving $k_0$${\bf k}$. The electric field, given by ${\bf E} = -{\bm \nabla}A_0$, reads
\begin{eqnarray} \label{eq_E_field}
{\bf E}_{\rm CFJ}({\bf x}) & = & \frac{1}{4\pi r^2} \Big\{ \left[  \left( {\bf k}\cdot{\bm \mu} \right) + 3 \left(  {\bf k}\cdot\hat{{\bf r}} \right) \left(  {\bm \mu} \cdot\hat{{\bf r}} \right)  \right] \hat{{\bf r}}  - \left(  {\bf k}\cdot\hat{{\bf r}} \right){\bm \mu} -  \left(  {\bm \mu} \cdot\hat{{\bf r}} \right){\bf k}  \Big\} \nonumber \\
& - & \frac{k_0}{2\pi r} \Big\{ \left(  {\bf k} \times {\bm \mu} \right)
-\left[ \hat{{\bf r}}\cdot \left(  {\bf k} \times {\bm \mu} \right) \right] \hat{{\bf r}} \Big\} \, .
\end{eqnarray}

%%%%%%%%%%%%%%%%%%%%%%%%%%%%%%%%%%%%%%%%%%%%%%%%%%
%%\begin{thebibliography}{99}

\bibliographystyle{unsrt}
\bibliography{refs}{}

%\end{thebibliography}
\end{document}